\documentclass[review,3p]{elsarticle}

\usepackage{comment}
\usepackage{wrapfig}
\usepackage{graphicx}
\usepackage{subcaption}
\usepackage{svg}
\usepackage{algorithm, algorithmic}
\usepackage{hyperref}
\usepackage{amssymb, amsmath, array,multirow}
\usepackage{makecell}
\usepackage{float}
\usepackage{placeins}
\usepackage[normalem]{ulem}
\usepackage{natbib} 
\usepackage{colortbl}
\usepackage{multirow}
\usepackage{hhline}
\usepackage{enumitem}

\begin{document}

\begin{frontmatter}
\title{Joint Estimation of Sea State and Vessel Parameters Using a Mass-Spring-Damper Equivalence Model}
\author[1]{Ranjeet K. Tiwari}
\ead{ranjeet.tiwari@adelaide.edu.au}
\author[2]{Daniel Sgarioto}
\ead{daniel.sgarioto@defence.gov.au}
\author[2] {Peter Graham}
\ead{peter.graham@defence.gov.au}
\author[2]{Alexei Skvortsov}
\ead{alexei.skvortsov@defence.gov.au}
\author[2]{Sanjeev Arulampalam}
\ead{sanjeev.arulampalam@defence.gov.au}
\author[1]{Damith C. Ranasinghe\corref{cor1}}
\ead{damith.ranasinghe@adelaide.edu.au}
%
\address[1]{School of Computer and Mathematical Science, The University of Adelaide, Adelaide, SA 5005, Australia}
\address[2]{Platforms Division,  Defence Science and Technology Group,  Melbourne, Vic, Australia}
\begin{abstract}
Real-time sea state estimation is vital for applications like shipbuilding and maritime safety. Traditional methods rely on accurate wave-vessel transfer functions to estimate wave spectra from onboard sensors. In contrast, our approach jointly estimates sea state and vessel parameters without needing prior transfer function knowledge, which may be unavailable or variable. We model the wave-vessel system using pseudo mass-spring-dampers and develop a dynamic model for the system. This method allows for recursive modeling of wave excitation as a  time-varying input, relaxing prior works' assumption of a constant input. We derive statistically consistent process noise covariance and implement a square root cubature Kalman filter for sensor data fusion. Further, we derive the Posterior Cramer-Rao lower bound to evaluate estimator performance. Extensive Monte Carlo simulations and data from a high-fidelity validated simulator confirm that the estimated wave spectrum matches methods assuming complete transfer function knowledge.

\end{abstract}

%
%

\end{frontmatter}


\section{Introduction}
\label{Intro}
Understanding sea waves and their interaction with a vessel is crucial for many maritime applications. Well-known examples include predicting arrival times, optimising mission paths, reducing hull fatigue, among  many others~\cite{jensen2001load, montazeri2016estimation, nielsen2020estimation, nielsen2019sea, pascoal2017estimation,jensen2004estimation}. Conventionally, wave buoys are used to measure wave properties. Increasing coverage of the oceans with wave buoys demands building new observation systems and their maintenance. The task is costly and information from dispersed wave buoys are not always accurate since they do not provide complete information about a particular vessel location. This necessitates exploring alternative options to better understand the sea state at a vessel's location. An attractive method is to employ sensors onboard a vessel to estimate wave properties, effectively treating the vessel as a wave buoy~\cite{nielsen2007response}.

Previous sea state estimation methods from ship sensor data introduced both parametric and non-parametric formulations for spectral estimation in the frequency domain~\cite{maeda2001estimation, masuda2001development, nielsen2006estimations, nielsen2007response, nielsen2008introducing, pascoal2005ocean, pascoal2008non}. The main limitation of these approaches is that their outputs are not available at short intervals in an online sense, and therefore are not suited to track gradual changes in the sea state with frequent updates as desired in real-time operational settings~~\cite{pascoal2009kalman}. Although recent advances have improved computational efficiency~[\cite{brodtkorb2023automatic, nielsen2022parameterised}, methods for obtaining the estimate at shorter intervals without full knowledge of the wave-vessel transfer functions remain an area of active interest.
 Consequently, researchers investigated methods for online estimation using onboard sensor measurements~\cite{pascoal2009kalman, pascoal2017estimation, kim2019real, kim2023real, andreas}.  Pioneering work in \cite{pascoal2009kalman} introduced a Kalman filter formulation for online estimation of wave properties, validated with synthesised datasets.  
Follow-on research extended the method to account for the motion of a moving ship 
\cite{pascoal2017estimation}. Later efforts aimed to refine these methods. For example, \cite{kim2019real} introduced a Wiener component to the transfer function in the Kalman filtering approach, aiming to mitigate overestimation at higher frequencies. The study in~\cite{kim2023real} further extended the inverse estimation approach to multi-directional sea waves, where both directions and frequencies are discretized to estimate the sea wave spectrum. These online estimation methods directly consider the elevation of components of a sea wave as part of the system's state. This formulation leads to a linear problem; however, its solution requires prior knowledge of the relationship between wave elevation and the specific vessel’s response to impacting sea waves modeled by a transfer function.

A key challenge with the existing online estimation problem formulations stems from the need to obtain an accurate transfer function for a vessel~\cite{montazeri2016estimation}.  Accurate knowledge of the transfer function, while often available in nominal form for modern vessels, may still be uncertain in operational settings due to its sensitivity to loading-dependent parameters such as draught, waterline breadth, centre of gravity, forward speed, and encounter conditions, as well as modelling assumptions and environmental variability~\cite{nielsen2019sea, brodtkorb2023automatic, mounet2024data,nielsen2022parameterised, van2023joint}. 
Consequently, our prior work in~\cite{van2023joint} proposed a new approach based on a sea-vessel system analogous to a pseudo mass-spring-damper model~\cite{gillijns2007unbiased, ma2004inverse, naets2015online} characterising the interaction between wave impact and vessel response. By comparing the dynamics of a mass-spring-damper system with known wave-vessel dynamics, the method derives parameters such as pseudo mass and pseudo damping coefficient in scenarios where vessel parameters are unknown. Notably, the approach offers a joint estimation of states, unknown wave excitations, and vessel parameters, thereby mitigating the need for accurate knowledge of the transfer function. However, the preliminary investigation focused on a feasibility demonstration with a \textit{single}-frequency wave impact, and the method could not account for the sea wave spectra encountered in practical settings, in both data acquisition and wave estimation.
 
In this work, we develop a method for online sea wave estimation for sea wave spectra that is resilient to incomplete knowledge of the transfer function parameters and to dynamic changes in the vessel’s physical characteristics, such as mass and breadth. We propose a \textit{system} of mass–spring–damper models, along with a wave–buoy equivalence, to formulate an input–state–parameter estimation problem using noisy measurements from onboard sensors, where sensor measurements and their {derived quantities} are treated as system states.
In this approach, the wave elevation is not directly considered as a state; instead, the wave excitation impacting the vessel—serving as the model input—is estimated jointly with the states and parameters. Estimating the instantaneous wave excitation then enables the estimation of key sea wave parameters of interest, such as the wave spectrum, significant wave height, and zero up-crossing period. In particular:

\begin{itemize}[itemsep=1pt,parsep=1pt,topsep=1pt,labelindent=0pt,leftmargin=5mm]
    \item  By expressing irregular sea waves in terms of regular waves (harmonics) using the superposition principle, we can employ the model of regular wave-vessel impact with a mass-spring-damper as the foundational element for constructing our system model for irregular sea waves. 
    This system model enhances computational feasibility and offers clear physical insights into the irregular wave-vessel interaction.
    \item We introduce dynamics to the process model, which can admit unknown vessel parameters and input wave excitation in its augmented state vector. 
     In prior work~\cite{van2023joint,pascoal2009kalman} the instantaneous wave elevation was assumed constant within the process model, with temporal variation captured indirectly via process noise. In contrast, we model the time evolution of the wave excitation through recursive state dynamics, thereby relaxing the assumption of time-invariant instantaneous wave elevation.
    \item Importantly, we compute the statistics of process noise, critical for implementing a Bayesian estimator, facilitated by the dynamic expression of the process model.
    \item We formulate an algorithm for the joint estimation of sea state and vessel parameters using the square-root cubature Kalman filter (SRCKF) for reconstructing the unknown irregular wave excitations along with vessel parameters. 
    \item To theoretically analyse the performance of the estimator implemented in this formulation, we derive the Posterior Cramer-Rao lower bound (PCRLB) for the irregular wave excitation. 
\end{itemize}

The implementation of our SRCKF-based measurement fusion approach yields improved estimation of vessel parameters and, consequently, a more accurate reconstruction of wave properties, comparable to a method that assumes complete knowledge of the transfer function.

The remainder of the paper is organised as follows. Section~\ref{sec:background} presents the necessary background on the Bayesian estimation framework, the nonlinear filtering approach adopted in this work, and the corresponding posterior error bound. Section~\ref{srckf} formulates the irregular wave model and integrates it with the regular wave–vessel interaction model to construct the complete wave–vessel dynamical system. Then, we describe the proposed nonlinear estimation algorithm and discuss the associated noise statistics. Next, we derive the posterior error bound for the estimator in Section~\ref{sec:bound}. Section~\ref{sec:exp} presents two experimental studies to validate the proposed approach: (i) controlled synthetic experiments with full ground-truth access for detailed spectrum and parameter analysis, and (ii) validation using a high-fidelity wave–vessel simulation framework benchmarked against towing-tank experiments.

\section{Background}\label{sec:background}
Given the nonlinearity and high dimensionality of the system model in the problem, we provide a brief background on Bayesian estimation in the context of nonlinear systems, as well as the performance bounds of estimators relevant to our work. We begin with an introduction to the notational aspects of our work. 

\subsection{Notations}\label{sec:notations}

Upright capital letters (e.g., $\mathrm{X}, \mathrm{Y}$) denote matrices, upright bold letters (e.g., $\mathbf{x}, \mathbf{y}$) denote vectors, and italic letters (e.g., $x, y$) denote scalar components.
Superscript $(\cdot)^s$ is used for variables representing the state space of the mass–spring–damper system, where $(\cdot)(t)$ and $(\cdot)_k$ denote the continuous- and discrete-time domains, respectively. The notation $(\cdot)_k^r$ represents variables from a regular system, where the unknown wave excitation is augmented with the displacement and velocity of the analogous mass–spring–damper system; whereas $(\cdot)_k^i$ denotes variables associated with the irregular system obtained by concatenating all regular components.
The notations $\hat{(\cdot)}_{k|k-1}$ and $\hat{(\cdot)}_{k|k}$ represent the predicted and posterior estimates of a variable $(\cdot)_k$, respectively.
The operator $\text{diag}(\cdot)$ denotes a diagonal matrix formed from the elements of $(\cdot)$, $\text{expm}(\cdot)$ denotes the matrix exponential, and $\text{blkdiag}(\cdot)$ denotes a block diagonal matrix formed from the matrices in $(\cdot)$.

\subsection{Bayesian Estimation of Nonlinear Systems}\label{sec:Bayes frame}
Suppose $\mathbf{x}_k$ is an unobserved Markov process at time step $k$ with a process model, $f(\mathbf{x}_{k-1}, \mathbf{q}_{k-1})$. Similarly, $\mathbf{y}_k$ is the available noisy information, which is independent of the previous information sequences, conditioned on $\mathbf{x}_k$, and it is related to the unobserved state by $g(\mathbf{x}_k, \mathbf{v}_k)$. Here, $\mathbf{q}_{k-1}$ and $\mathbf{v}_k$ are process and measurement noises, which are independent and mutually uncorrelated sequences with known distributions.  Then, in general, an estimator for the unobserved state $\mathbf{x}_k$ can be developed by following the two-step Bayesian framework of predict and update procedures where the predicted density, $P(\mathbf{x}_{k}|\mathbf{y}_{1:k-1})$, is given by Chapman-Kolmogorov integral:\\
 \begin{equation}\label{prior}
     P(\mathbf{x}_{k}|\mathbf{y}_{1:k-1}) = \int P(\mathbf{x}_k|\mathbf{x}_{k-1})P(\mathbf{x}_{k-1}|\mathbf{y}_{1:k-1})\mathrm{d}\mathbf{x}_{k-1},
 \end{equation}
 and by using the Bayes law, the posterior density, $P(\mathbf{x}_k|\mathbf{y}_{1:k})$, is computed by the update step: \begin{equation}\label{post}
 P(\mathbf{x}_k|\mathbf{y}_{1:k}) = \dfrac{P(\mathbf{y}_k|\mathbf{x}_k)P(\mathbf{x}_{k}|\mathbf{y}_{1:k-1})}{P(\mathbf{y}_k|\mathbf{y}_{1:k-1})}.
 \end{equation}
 
 Here, the transitional density, $P(\mathbf{x}_k|\mathbf{x}_{k-1})$, is constructed with the help of process dynamics and its noise statistics,  whereas the likelihood density, $P(\mathbf{y}_k|\mathbf{x}_k)$, is derived by using the measurement model and the statistics of measurement noise. Now, if the system model is linear and the statistics of its noises follow the Gaussian distribution, the likelihood and transitional densities remain a Gaussian, and the predicted density and the normalizing constant, $P(\mathbf{y}_k|\mathbf{y}_{1:k-1})$, can be computed analytically (assuming the prior is also Gaussian). In this situation, the posterior density has a closed-form solution, which is also optimal; an example of such an estimator is the Kalman filter \cite{kalman1960}, used in existing sea state estimation work. 

 In contrast, if the noise statistics follow a Gaussian distribution but the system model is nonlinear, the likelihood and transition densities become non-Gaussian, and despite the prior being Gaussian, a closed-form solution for the posterior density is not tractable. In such cases, broadly, two approaches are computationally efficient: i)~ linearise the system model piecewise at each time step,
 known as extended Kalman filtering (EKF); and ii)~instead of linearising the models, effectively changing the system, approximate the integrals with a set of weighted deterministic points. If the prior is Gaussian, the integrands of these integrals are a nonlinear function (from the system model) and a Gaussian density (prior), which enables the use of several approximation techniques for multidimensional integrals under the Gaussian environment, such as spherical-radial cubature rule \cite{arasaratnam2009cubature}, cubature and quadrature rule \cite{bhaumik2013cubature}, Hermite expansion \cite{arasaratnam2007discrete},  and orthogonal chaos expansion \cite{kumar2023polynomial}, among others. This approximation is generally employed to compute the predicted densities for state and measurement (normalising constant), and subsequently, they are assumed to be Gaussian for further computation of the posterior.

Instead of approximating with a set of deterministic points, these Gaussian integrals can also be computed with a set of stochastically chosen points that are weighted according to their likelihood of representing the posterior. This sequential Monte Carlo (SMC)~\cite{Gordon1993,doucet2009tutorial} method is applied in the case of highly nonlinear systems or those with non-Gaussian noise. Alternatively, sum of Gaussian mixtures approaches are also proposed in the literature for dealing with non-Gaussian problems~\cite{leong2013gaussian}. In our work, we consider a formulation of a cubature Kalman filter (CKF) for the resulting nonlinear system.

\subsection{Cubature Kalman Filter}\label{sec:ckf}
{For the underlying high-dimensional nonlinear estimation problem, we adopt a CKF~\cite{arasaratnam2009cubature} instead of an EKF.  
Unlike an EKF formulation, which relies only on first-order terms in the Taylor expansion, a CKF captures up to third-order terms~\cite{zhao2016performance} and facilitates a simpler implementation. In addition, a CKF serves as a more computationally efficient alternative to Sequential Monte Carlo (SMC) methods, which become prohibitively expensive for high-dimensional systems. 
A brief overview of a CKF is provided here to support the discussion of the algorithm developed for the sea-vessel parameter estimation problem.}

The CKF assumes that the resulting density after the approximation of the integral of the product of a Gaussian prior and a nonlinear function remains Gaussian. Hence, the predictive, from \eqref{prior}, and posterior, from \eqref{post}, densities remain Gaussian in a cycle over the time steps. A Gaussian density can be characterized by its mean and covariance, and that of $P(\mathbf{x}_k|\mathbf{y}_{1:k-1})$ from \eqref{prior} for the underlying system can be given as
 \begin{equation}
 \begin{split}
     \hat{\mathbf{x}}_{k|k-1} & = \int f(\mathbf{x}_{k-1}, q_{k-1}) \mathcal{N}(\mathbf{x}_{k-1}; \hat{\mathbf{x}}_{k-1|k-1}, \Sigma_{k-1|k-1}) \mathrm{d}\mathbf{x}_{k-1},\\ \label{pred_mean1}
     \Sigma_{k|k-1} & = \int (f(\mathbf{x}_{k-1}, q_{k-1})-\hat{\mathbf{x}}_{k|k-1})(f(\mathbf{x}_{k-1}, q_{k-1})-\hat{\mathbf{x}}_{k|k-1})^\top \mathcal{N}(\mathbf{x}_{k-1}; \hat{\mathbf{x}}_{k-1|k-1}, \Sigma_{k-1|k-1}) \mathrm{d}\mathbf{x}_{k-1}, 
     \end{split}
 \end{equation}
 where $ \hat{\mathbf{x}}_{k-1|k-1}$ and $\Sigma_{k-1|k-1}$ are the mean and covariance of the prior, $P(\mathbf{x}_{k-1}|\mathbf{y}_{1:k-1})$, respectively, and they are assumed to be known at time step $k$. Now, using a set of $2n_{\mathbf{x}}$ equi-weighted deterministic points from the spherical-radial rules \cite{arasaratnam2009cubature}, the above two moments of the predicted density can be approximated. We formulate the predicted density for the system model in this study in Section~\ref{sec:joint estimation}.

 A two-step Bayesian filter first uses the prior density and process model to construct the predictive density, which is also known as the time-update and given in \eqref{pred_mean1} for a generalized system. Subsequently, this predicted density is corrected in the second step using the current measurement, $\mathbf{y}_k$.  The expression for the posterior density in \eqref{post} can be rewritten as
 \begin{equation}
     P(\mathbf{x}_k|\mathbf{y}_{1:k}) = \dfrac{P(\mathbf{y}_k,\mathbf{x}_k|\mathbf{y}_{1:k-1})}{P(\mathbf{y}_k|\mathbf{y}_{1:k-1})}.
 \end{equation}
 Again, we expand the joint density, $P(\mathbf{y}_k,\mathbf{x}_k|\mathbf{y}_{1:k-1})$, and the measurement predictive density, $P(\mathbf{y}_k|\mathbf{y}_{1:k-1})$ by using the assumed multivariate Gaussian structures and {rearrange them into a standard Gaussian distribution form} \cite{wang2013gaussian}. Consequently, we obtain a Gaussian posterior with mean and covariance as below:
 \begin{align}
     \hat{\mathbf{x}}_{k|k} &= \hat{\mathbf{x}}_{k|k-1} +\mathrm{L}_k(\mathbf{y}_k-\hat{\mathbf{y}}_{k|k-1}),\label{post_mean}\\
     \Sigma_{k|k} &= \Sigma_{k|k-1}-\mathrm{L}_k\Sigma_{k|k-1}^{\mathbf{yy}}L_k^\top,\label{post_cov}
 \end{align}
 where \  
     $\mathrm{L}_k = \Sigma_{k|k-1}^{\mathbf{xy}}{\Sigma_{k|k-1}^{{\mathbf{yy}}^{-1}}}.$
 Here, the predicted measurement $\hat{\mathbf{y}}_{k|k-1}$, innovation covariance $\Sigma_{k|k-1}^{\mathbf{yy}}$, and cross-covariance $\Sigma_{k|k-1}^{\mathbf{xy}}$ are computed over the predictive density, $P(\mathbf{x}_k|\mathbf{y}_{1:k-1})$, as below:
 \begin{equation*}
 \begin{split}
     \hat{\mathbf{y}}_{k|k-1}&= \int g(\mathbf{x}_k, v_k) \mathcal{N}(\mathbf{x}_k; \hat{\mathbf{x}}_{k|k-1}, \Sigma_{k|k-1})\mathrm{d}\mathbf{x}_k,\\
     \Sigma_{k|k-1}^{\mathbf{yy}} & = \int (g(\mathbf{x}_k, v_k)-\hat{\mathbf{y}}_{k|k-1})(g(\mathbf{x}_k, v_k)-\hat{\mathbf{y}}_{k|k-1})^\top \mathcal{N}(\mathbf{x}_k; \hat{\mathbf{x}}_{k|k-1}, \Sigma_{k|k-1})\mathrm{d}\mathbf{x}_k,\\
     \Sigma_{k|k-1}^{\mathbf{xy}} & = \int (\mathbf{x}_k-\hat{\mathbf{x}}_{k|k-1})(g(\mathbf{x}_k, v_k)-\hat{\mathbf{y}}_{k|k-1})^\top\mathcal{N}(\mathbf{x}_k; \hat{\mathbf{x}}_{k|k-1}, \Sigma_{k|k-1})\mathrm{d}\mathbf{x}_k.
     \end{split}
 \end{equation*}
 Again, the integrals are approximated with cubature points. We describe the approximations for the wave-vessel system in Section~\ref{sec:joint estimation}.

 One requirement for implementing the CKF algorithm in its standard form is deriving the square-root matrix of the covariance---see~\eqref{prior_x} and \eqref{predicted_x}---which calls for factorization methods that can convert a symmetric and positive definite matrix into the product of lower and upper triangular matrices. Often a software that facilitates implementing the algorithm loses the positive definiteness of the covariance matrix due to limited precision (number of significant figures) and makes factorization numerically infeasible. To improve the numerical stability of CKF, its square-root version was developed \cite{arasaratnam2009cubature}, which removes the need for factorization of matrices. 

\subsection{Performance Bounds}\label{sec:performance bounds}
Ideally, when designing an estimator for a system, its performance with respect to  the minimum possible estimation error is valuable for understanding the goodness of the estimator. This lower bound on estimation error, specific to a given system model and available measurement data, serves as a benchmark for assessing the estimator's accuracy. Importantly, for a discrete-time nonlinear system  where an estimator is suboptimal (such as that confronted in our work), it is often difficult to understand the achievable performance, especially in a relatively new application where a  benchmark for the problem is lacking in the literature. Therefore a theoretical bound provides valuable insights.

Several lower bounds exist in the literature for different system types~\cite{bobrovsky1975lower, galdos1980cramer, kerr1989status, tichavsky1995posterior}; most generalise the Cramér-Rao lower bound (CRLB)~\cite{kerr1989status}. Initially, the CRLB was introduced as a scalar measure for time-invariant systems with unbiased estimators. The Van Trees version of the CRLB~\cite{van2004detection}, also known as the posterior CRLB (PCRLB), is widely used to compute lower bounds for estimated time-varying parameters in nonlinear and potentially non-Gaussian systems~\cite{tichavsky1995posterior, tichavsky1998posterior}. Notably, the estimator we develop for the state in our problem is based on a nonlinear system. Consequently, we derive the PCRLB for our estimator to analyse and understand its performance.
 
 \section{Joint Estimation of Input-State-Parameters}\label{srckf} 

Our objective is to estimate the properties of irregular sea waves, such as the wave spectrum, significant wave height, and zero up-crossing period, using measurements from onboard sensors without prior knowledge of the
vessel’s parameters---specifically, the system parameters, the vessel’s breadth and draught. First, we propose an irregular wave–vessel system model used to estimate the irregular wave excitations and the unknown vessel parameters with measurements of the vessel's vertical motions. Second, a sliding-window-based fast Fourier transform (FFT) is applied to the estimated instantaneous excitations, which computes the wave spectrum and associated wave properties.

To construct the irregular wave–vessel system model, we employ a sea keeping analysis and its analogue mass–spring–damper dynamics. The conceptual equivalence between a mass–spring–damper model and wave–vessel interaction, as used in~\cite{van2023joint}, is valid for a regular wave (i.e., a sinusoid with a single frequency component). In this representation, each regular vertical wave excitation and the dynamics of vessel motion is modeled by a mass–spring–damper system, connected through the shared vessel parameters. Building upon this concept, we extend the model to represent a full sea wave spectrum, which is more representative of practical sea conditions.
Specifically, an irregular wave is expressed as a superposition of multiple regular components with different frequencies, amplitudes, and phases. Consequently, the overall wave–vessel interaction is represented by a system of simultaneous mass–spring–damper models, coupled through the common vessel. Using this framework
and onboard sensor measurements, we aim to jointly estimate the states, the heave excitation, the pitch excitation, and the unknown system parameters, namely waterline breadth and draught.

 \subsection{Wave-Vessel System}\label{sys_mod}
\begin{figure}[h!]
	\centering
\begin{subfigure}{0.45\textwidth}
	\includegraphics[width=\textwidth]{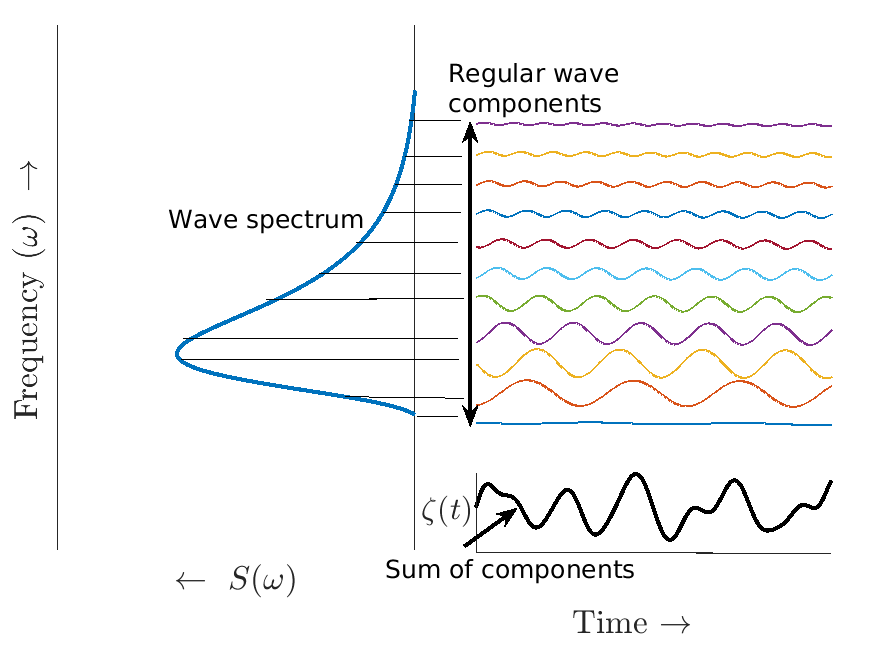}
	\caption{}
	\label{fig1}
\end{subfigure}
\begin{subfigure}{0.41\textwidth}
	\includegraphics[width=\textwidth]{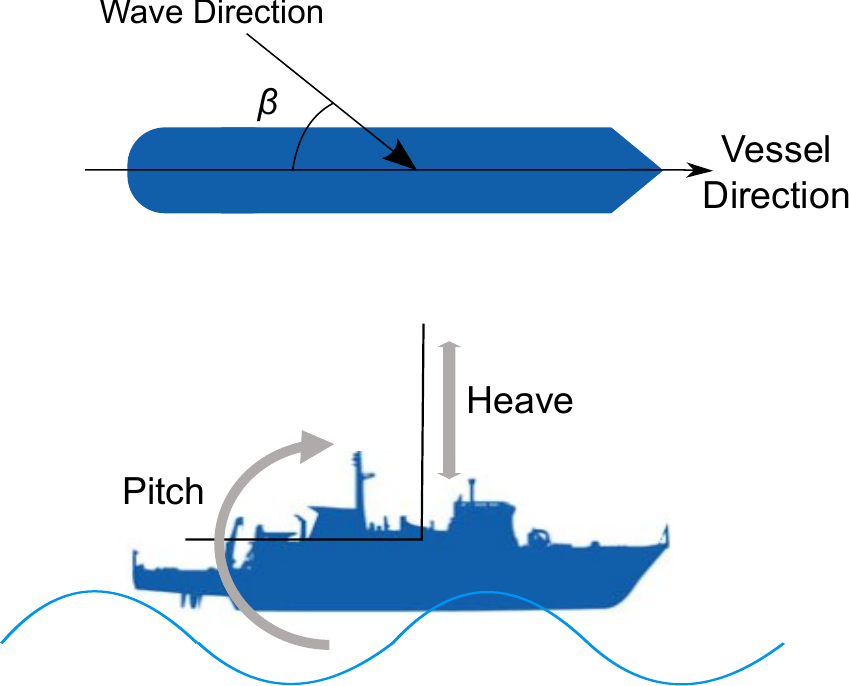}
	\caption{}
	\label{fig2}
\end{subfigure}

\vspace{-4mm}
\caption{(a) Illustration of an irregular wave form and its constituent harmonics in time and frequency domains. (b) Sea wave-induced vessel response captured by the heave and pitch motions of a sea vessel. }
\vspace{-2mm}
\end{figure}

We can express an irregular sea wave in the time domain as the sum of $N$ regular (harmonic) waves. Then, a state-space expression for the wave elevation of an irregular and {long-crested} wave under the assumption of zero forward speed can be given as
 \begin{equation}\label{irr_wave}
     \zeta(t) = \sum_{n=1}^N \tilde{\zeta}_n {\sin} (\omega_nt + \epsilon_n),
 \end{equation}
 where $\tilde{\zeta}_n$ is the amplitude of $n$th wave component with frequency $\omega_n$ and phase difference $\epsilon_n$. In this work, we restrict our attention to long-crested, unidirectional waves \cite{micaela,kim2019real}. For a given frequency range, the amplitudes of the constituent waves are obtained from a prescribed wave spectrum; however, their phases ($\epsilon_n$) are assigned randomly from a uniform distribution over $[0, \ 2\pi]$. A descriptive illustration of an irregular and {unidirectional} wave showing its harmonic components in the time and frequency domains is shown in Fig.~\ref{fig1}. Next, we model the process dynamics and measurements for a single regular component of \eqref{irr_wave} and then extend it for an irregular sea wave. 
 
\subsubsection{Process and Measurement Model} 
We begin by describing the analogue between regular wave induced heave and pitch motions of a wave-vessel system and the state dynamics of a \textit{pseudo} mass–spring–damper under an external excitation formulated in continuous-time time domain. This continuous-time model is then discretised, and the unknown excitation is augmented with the state vector, yielding the regular state model of the proposed method, denoted by $(\cdot)^r$. 
Subsequently, constructing an irregular sea wave from regular components and exploiting the linearity of the mass–spring–damper equivalence, a regular state model is defined for each component of the irregular wave excitation, which are then combined into a single irregular process model (denoted by $(\cdot)^i$) for the sea wave–vessel interaction. The resultant linear process model is further augmented with the unknown vessel parameters, resulting in the complete system model. This model is nonlinear, owing to the dependence of the states on the unknown parameters.

We begin with the continuous-time modeling of wave excitation and vessel response. 
This work adopts the simplified seakeeping analysis of a regular wave–vessel system presented in \cite{jensen2004estimation}, where the vessel is assumed to be box-shaped and uniformly loaded. The governing dynamics are given as follows:
 \begin{align}\label{eq:nelson-heave}
\dfrac{2T}{g}\ddot{\tau} + \dfrac{A_v^2}{k_wB\alpha^3\omega}\dot{\tau}+\tau &= \tilde{\zeta}P_\tau\sin (\bar{\omega}t +\phi_\tau)=p_\tau(t)\\ 
\dfrac{2T}{g}\ddot{\theta} + \dfrac{A_v^2}{k_wB\alpha^3\omega}\dot{\theta}+\theta &= \tilde{\zeta}P_{\theta}\sin (\bar{\omega}t +\phi_{\theta})=p_{\theta}(t),\label{eq:nelson-pitch}
 \end{align}
 where heave $\tau$ and pitch $\theta$ are considered as two single-degree-of-freedom (DoF) vessel responses. Then, since the coupling between these two vertical motions can be neglected~\cite{jensen2004estimation}, the phase difference between the corresponding wave excitations, $\phi_\tau$ and $\phi_{\theta}$, is $\pi/2$. Notably, in long waves, the heave response is synchronised with the wave motion, whereas the pitch response follows the wave slope and remains $\pi/2$ out of phase with the wave motion.  Here,  $k_w=\omega^2/g$ is the wave number, $g=9.8$ $\mathrm{m/s^2}$ is the acceleration due to gravity, $\tilde{\zeta}$ is the wave amplitude, $B$ and $T$ are the waterline breadth and draught of the vessel, $\omega$ and $\bar{\omega}$ are the incident wave and encountered wave frequencies, respectively, and $p_\tau$ is the input heave excitation with amplitude $\tilde{\zeta}P_\tau$ in meters and $p_\theta(t)$ is the input pitch excitation with amplitude $\tilde{\zeta}P_{\theta}$ in radians. $P_\tau$ and $P_\theta$ (defined in \eqref{force_w} and \eqref{force_theta}) are the forcing functions---attributed to the geometry of the vessel---for the wave excitations, $p_\tau(t)$ and $p_\theta(t)$, respectively.

 The wave frequency experienced on the vessel (called encountered frequency) and the incident sea wave frequency are related by Doppler shift, and are expressed as
 \begin{equation}\label{doppler}
 \bar{\omega} = \omega - k_wV \cos(\beta),
 \end{equation}
 where $\beta$ is the relative heading between vessel motion and sea wave propagation, and $V$ is the vessel forward speed. {Now, if the vessel has a length $L$, the remaining parameters describing the vessel response under wave excitation can be defined following \cite{jensen2004estimation} as
$\alpha = 1 - V\sqrt{\dfrac{k_w}{g}} \cos(\beta)$, $A_v = 2\sin (\dfrac{k_wB\alpha^2}{2})\exp (-k_wT\alpha^2)$,
 $k_e = |k_w\cos(\beta)|$,
 $\kappa = \exp(-k_wT),$
 $\psi = \sqrt{(1-k_wT)^2 +(\dfrac{A_v^2}{k_wB\alpha^3})^2}$, and the forcing functions are expressed as}
 \begin{equation}\label{force_w}
 P_\tau = 2\kappa \psi \dfrac{\sin(\dfrac{k_eL}{2})}{(k_eL)},
 \end{equation}
 \begin{equation}\label{force_theta}
 P_{\theta} = 24\kappa \psi \dfrac{\sin(\dfrac{k_eL}{2})-\dfrac{1}{2}k_eL\cos(\dfrac{k_eL}{2})}{k_e^2L^3}.
 \end{equation}

Subsequently, to model the {regular} wave-vessel dynamics described in \eqref{eq:nelson-heave} and \eqref{eq:nelson-pitch}, we consider two pseudo mass-spring-damper systems with {decoupled} single DoF motions under the influence of two wave excitations induced by a regular sea wave. Considering the standard dynamics for a single DoF mass-spring-damper system, we can formulate a model for the vessel experiencing heave or pitch motion as:
  \begin{equation}\label{msd}
      M(\boldsymbol{\eta})\ddot{x}(t) +C(\boldsymbol{\eta})\dot{x}(t) +x(t) = p(t).
  \end{equation}
  {Here, $x(t)$ and $p(t)$ denote the state and excitation input of the system---either a heave excitation $p_\tau(t)$ or a pitch excitation $p_{\theta}(t)$---with pseudo-mass coefficient $M$ and pseudo-damping coefficient $C$, both expressed as functions of the parameter vector $\boldsymbol{\eta}$. The dynamics in~\eqref{msd} are normalised by the pseudo-stiffness coefficient.
By comparing the dynamics in~\eqref{eq:nelson-heave} and~\eqref{eq:nelson-pitch} with that in~\eqref{msd}, the expressions for the equivalent pseudo-mass and pseudo-damping coefficients in~(13) can be expressed as follows:}
  \begin{align}
      M(\boldsymbol{\eta})&= \dfrac{2T}{g}, \label{M}\\
      C(\boldsymbol{\eta})&= \dfrac{g{A_v}(\boldsymbol{\eta})^2}{B{\omega}^3\alpha^3},\label{C}
  \end{align}
  {where the sectional damping approximation, $A_v(\boldsymbol{\eta})$, the Doppler shift, $\alpha$ and the parameters, $\boldsymbol{\eta}=[B, T]^\top$ are defined as discussed earlier. 
  } Notably, in a generalised seakeeping analysis, where a vessel is considered as a band of strips, a higher DoF mass-spring-damper system \cite{van2020computationally} should be used to effectively represent the interaction among different strips of the vessel under the influence of a wave force.

  Now, {we define the regular displacement (heave or pitch) and velocity as the states of the analogous mass-spring-damper system defined in \eqref{msd}}, ${\mathbf{x}^s(t)}=[x(t), \dot{x}(t)]^\top \in \Re^{n_{x^s}}$, the state-space expressions for the system are 
  \begin{align}\label{process_c}
     { \dot{\mathbf{x}}^s(t)} &= {\mathrm{A}^s(\boldsymbol{\eta}, t)\mathbf{x}^s(t) + \mathbf{B}^s(\boldsymbol{\eta}, t)}p(t),\\
     { \mathbf{y}^s(t)} &= {\mathrm{G}^s(\boldsymbol{\eta}, t)\mathbf{x}^s(t) + \mathbf{J}^s(\boldsymbol{\eta},t)}p(t),\label{meas_c}
  \end{align}
  where 
     $ {\mathrm{A}^s(\boldsymbol{\eta}, t)} = \begin{bmatrix} 0 &1\\-1/M(\boldsymbol{\eta}) & -C(\boldsymbol{\eta})/M(\boldsymbol{\eta}) \end{bmatrix}~~\text{and}~~{\mathbf{B}^s(\boldsymbol{\eta}, t) } = \begin{bmatrix}0\\  1/M(\boldsymbol{\eta})\end{bmatrix}.$
  In this formulation, three types of measurements, namely position, $y(t)$, velocity, $\dot{y}(t)$, and acceleration, $\ddot{y}(t)$, are considered for the joint estimation of system input and parameters along with the states. If $\mathbf{y}^s(t) = [y(t), \dot{y}(t), \ddot{y}(t)]^\top$, then, 
  \begin{equation*}
      {\mathrm{G}^s(\boldsymbol{\eta}, t)} = \begin{bmatrix} 1&0\\0&1\\-1/M(\boldsymbol{\eta}) & -C(\boldsymbol{\eta})/M(\boldsymbol{\eta})\end{bmatrix}~~\text{and}~~{\mathbf{J}^s(\boldsymbol{\eta}, t) = \begin{bmatrix} 0\\0\\1/M(\boldsymbol{\eta}) \end{bmatrix}}.
  \end{equation*}
 To implement a recursive Bayesian estimator when measurements are available at discrete intervals, a discrete-time representation of the underlying system is preferred. For accuracy, we choose a first-order hold (FOH) discretisation over a zero-order hold (ZOH), despite its higher computational cost. Applying the FOH method~(p.757 of \cite{maes2019tracking}) with a sampling time of $T_s$, the continuous-time system in \eqref{process_c} and \eqref{meas_c} can be discretized as  
\begin{align}\label{sys_dicrete}
    \mathbf{x}_k^s &= \mathrm{A}_k^s(\boldsymbol{\eta})\mathbf{x}_{k-1}^s + \mathbf{B}^s_k(\boldsymbol{\eta})p_{k-1} + \mathbf{q}_{k-1}^s,\\
    \mathbf{y}^s_k   &= \mathrm{G}^s_k(\boldsymbol{\eta})\mathbf{x}_k^s + \mathbf{J}^s_k(\boldsymbol{\eta})p_k + \mathbf{r}_k,\label{discre_meas}
\end{align}
where $\mathbf{q}_{k-1}^s$ and $\mathbf{r}^s_k$ are process and measurement noise sequences, respectively. The additive process noise accounts for unmodeled hydrodynamic effects, simplifications in the wave–vessel interaction model, and discrepancies between the assumed dynamics and the actual vessel response. The measurement noise represents sensor-related uncertainties and other disturbances affecting the recorded heave and pitch measurements that are not explicitly captured in the deterministic measurement model. Both are assumed to be mutually uncorrelated, zero-mean Gaussian white noise with distributions  
$\mathbf{q}_k^s \sim \mathcal{N}(0, \mathrm{Q}_k^s)$ and $\mathbf{r}^s_k \sim \mathcal{N}(0, \mathrm{R}^s_k)$. Then,  
  {\begin{align*}
      \mathrm{A}^s_k(\boldsymbol{\eta}) &= \text{expm}(\mathrm{A}^s(\boldsymbol{\eta}, t)T_s),\
      \mathrm{B}^s_k(\boldsymbol{\eta})  = \dfrac{1}{T_s}\mathrm{A}^s(\boldsymbol{\eta}, t)^{-2}(\mathrm{A}^s_k(\boldsymbol{\eta})-\mathrm{I})^2\mathbf{B}^s(\boldsymbol{\eta}, t),\\
      \mathrm{G}^s_k(\boldsymbol{\eta})& = \mathrm{G}^s(\boldsymbol{\eta}, t),~\text{and}\
      \mathbf{J}^s_k(\boldsymbol{\eta}) = \mathbf{J}^s(\boldsymbol{\eta}, t) +\mathrm{G}^s_k(\boldsymbol{\eta})\left(\dfrac{\mathrm{A}^s(\boldsymbol{\eta}, t)^{-2}}{T_s}(\mathrm{A}^s_k(\boldsymbol{\eta})-\mathrm{I})-\mathrm{A}^s(\boldsymbol{\eta}, t)^{-1}\right)\mathbf{B}^s(\boldsymbol{\eta}, t).
  \end{align*}}
  Since we aim to jointly estimate the state-input-parameters using onboard sensors measurements, then we augment our state vector with the unknown wave excitation signal and define a state vector that completely describes the unobserved states of the regular system,  $\mathbf{x}^r_k = [\mathbf{x}_k^{s \top}, p_k]^\top$. Here, using \eqref{eq:nelson-heave} and  \eqref{eq:nelson-pitch}, the wave excitation can be expressed in the discrete-time domain as

  \vspace{-6mm}
  \begin{flushleft}
  \begin{equation}\label{reg_input}
      p_k = \tilde{\zeta}P\sin (\bar{\omega}kT_s +\phi).
  \end{equation}

  Further, to express the wave excitation recursively, expanding its value at the next time step and rearranging the right-hand side, we can write
 \vspace{-3mm}
  \begin{equation*}
      p_{k+1} = \tilde{\zeta}P\sin(\bar{\omega}kT_s +\phi)\cos(\bar{\omega}T_s) +\tilde{\zeta}P\cos(\bar{\omega}kT_s+\phi)\sin(\bar{\omega}T_s).
  \end{equation*} Substituting $p_k$ from \eqref{reg_input}, it can be expressed recursively as
  \begin{equation}\label{input_recur}
      p_{k+1} = p_k\left(\cos(\bar{\omega}T_s) +\dfrac{\sin(\bar{\omega}T_s)}{\tan (\bar{\omega}kT_s +\phi)}\right).
  \end{equation}
  
Then, using \eqref{sys_dicrete}, \eqref{discre_meas}, and \eqref{input_recur}, we have
 \vspace{-3mm}
  \begin{align}\label{reg_model}
      \mathbf{x}_k^r&= \mathrm{A}^r_k(\boldsymbol{\eta})\mathbf{x}_{k-1}^r  +\mathbf{q}_{k-1}^r,\\
      \mathbf{y}_k &= \mathrm{G}^r_k(\boldsymbol{\eta})\mathbf{x}_k^r +\mathbf{r}_k, \label{reg_meas}
  \end{align}
  \begin{align}
    \text{ where } \quad  \mathrm{A}^r_k(\boldsymbol{\eta}) =\begin{bmatrix} \mathrm{A}^s_k(\boldsymbol{\eta}) &\mathbf{B}^s_k(\boldsymbol{\eta})\\ \mathbf{0} & \left(\cos(\bar{\omega}T_s)+\dfrac{\sin(\bar{\omega}T_s)}{\tan (\bar{\omega}(k-1)T_s +\phi)}\right)\mathrm{I}_{n_p} \end{bmatrix},  \quad
      \mathrm{G}^r_k(\boldsymbol{\eta})= \begin{bmatrix} \mathrm{G}^s_k(\boldsymbol{\eta}) & \mathbf{J}^s_k(\boldsymbol{\eta}) \end{bmatrix}, \label{reg_transition}
  \end{align}
  \end{flushleft}
  and $\mathbf{q}_k^r \sim \mathcal{N}(\mathbf{0}, \mathrm{Q}_k^r)$ is a Gaussian sequence with zero mean and covariance $\mathrm{Q}_k^r = \text{blkdiag}(\mathrm{Q}_k^s, \mathrm{Q}_k^p)$,  such that $\mathrm{Q}_k^p \in \Re^{n_p \times n_p}$ represents the uncertainty in wave excitation modeling. Notably, for estimation purposes, the encountered frequency range must be chosen such that $\tan(\bar{\omega}kT_s + \phi)$ in \eqref{input_recur} does not evaluate to zero. However, owing to the random nature of $\phi$, it may become zero at certain time steps. At such instances, we replace the recursive expression of the regular excitation in the dynamics with a non-recursive update, given by $p_k = p_{k-1}$. 

Next, we use this wave-vessel system model for regular sea waves to design the model for the interaction between irregular wave impact and vessel response. Then, augment the resultant irregular state vector with unknown parameters to obtain the model for the proposed joint estimation under irregular wave impact.
  
\vspace{-2mm}
\begin{wrapfigure}{l}{0.48\textwidth}
	\centering
	\includegraphics[width=0.48\textwidth]{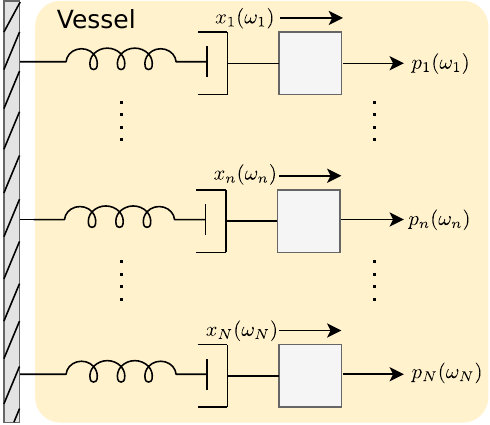}
	\caption{The wave-vessel interaction model using a set of $N$ interconnected mass-spring-damper blocks, each representing a regular wave-vessel interaction, where $\sum_n x_n$ is the vessel's collective response when impacted by an irregular wave excitation, $\sum_n p_n(\omega_n)$.}
	\label{fig_msd}
\vspace{-5mm}
\end{wrapfigure}

We assume that the irregular wave is composed of $N$ harmonics (regular sinusoidal waves) and each regular wave-vessel system behaves linearly, so that the principle of superposition can be applied for developing the irregular wave-vessel model using the model of its constituent regular systems. Figure~\ref{fig_msd} illustrates the resulting irregular system represented by a set of $N$ simultaneous mass-spring-damper systems, with each block representing a constituent wave-vessel interaction. This depiction visually captures how a vessel integrates the responses associated with individual components of an irregular wave.
Hence, using the expression in \eqref{irr_wave} and the model described in \eqref{reg_model} and \eqref{reg_meas}, we can write the state-space model as
 \begin{align}
 \mathbf{x}_k^i &= \mathrm{A}^i_k(\boldsymbol{\eta}) \mathbf{x}_{k-1}^i + \mathbf{q}_{k-1}^i,\label{irr_process}\\
 \mathbf{y}_k &= \mathrm{G}^i_k(\boldsymbol{\eta})\mathbf{x}_k^i + \mathbf{r}_k, \label{irr_meas}
 \end{align}
 where the state is defined as $\mathbf{x}_k^i = [\mathbf{x}_{1,k}^{r\top}, \cdots, \mathbf{x}_{N,k}^{r\top}]^\top$ and \ 
     $\mathrm{A}^i_k(\boldsymbol{\eta}) = \begin{bmatrix} \mathrm{A}^r_{1,k}(\boldsymbol{\eta}) &\mathrm{0}_{n_{x^r}\times n_{x^r}}&\cdots &\mathrm{0}_{n_{x^r}\times n_{x^r}}\\ 
     \mathrm{0}_{n_{x^r}\times n_{x^r}}& \mathrm{A}^r_{2,k}(\boldsymbol{\eta}) &\cdots&\mathrm{0}_{n_{x^r}\times n_{x^r}}\\
     \vdots& &\ddots \\
     \mathrm{0}_{n_{x^r}\times n_{x^r}}& \cdots & &\mathrm{A}^r_{N,k}(\boldsymbol{\eta}) \end{bmatrix},\label{irr_A}\\ 
     \mathrm{G}^i_k(\boldsymbol{\eta})  = \begin{bmatrix} \mathrm{G}^r_{1,k}(\boldsymbol{\eta})& \cdots & \mathrm{G}^r_{N,k}(\boldsymbol{\eta}) \end{bmatrix},$ \label{irr_G}
  and $\mathbf{q}_k^i \sim \mathcal{N}(\mathbf{0}, \mathrm{Q}_k^i)$. $\mathrm{Q}_k^i$ is constructed with corresponding regular components and $\mathrm{Q}_k^r$ in a similar fashion to $\mathrm{A}^i_k(\boldsymbol{\eta})$. The model described in \eqref{irr_process} and \eqref{irr_meas} governs the manner in which the state and input of the irregular wave-vessel system evolve over time. 
  
  To develop an estimator for the system,  we need to integrate the unknown vessel parameters into the unobserved system state. The parameter $\boldsymbol{\eta}$ is common to every constituent regular component of irregular wave dynamics, and does not need to be estimated separately for each harmonic. Hence, we add the unknown parameter, $\boldsymbol{\eta}$, to our irregular state-space representation and define the resulting augmented state \begin{equation}
      \mathbf{x}_k = [\mathbf{x}_k^{i^\top}, \boldsymbol{\eta}^\top]^\top.\label{final_state}
  \end{equation} The modified system model can then be expressed as 
  \begin{align}
      \mathbf{x}_k& = \mathrm{A}_k(\boldsymbol{\eta}) \mathbf{x}_{k-1} + \mathbf{q}_{k-1},\label{process_dyna}\\
      \mathbf{y}_k & = \mathrm{G}_k(\boldsymbol{\eta})\mathbf{x}_k + \mathbf{r}_k,\label{meas_model}
  \end{align}
  where 
     $\mathrm{A}_k(\boldsymbol{\eta})  = \begin{bmatrix} \mathrm{A}^i_k(\boldsymbol{\eta}) & \mathrm{0}_{n_{\mathbf{x}^i} \times n_{\boldsymbol{\eta}}}\\
      \mathrm{0}_{n_{\mathbf{x}^i}} \times n_{\boldsymbol{\eta}} & \mathrm{I}_{n_{\boldsymbol{\eta}}} \end{bmatrix},~~\mathrm{G}_k(\boldsymbol{\eta})  = \begin{bmatrix}\mathrm{G}^i_k(\boldsymbol{\eta})& \mathrm{0}_{n_{\mathbf{y}}} \times n_{\boldsymbol{\eta}}\end{bmatrix},$
  and the process noise, $\mathbf{q}_k \sim \mathcal{N}(\mathbf{0}, \mathrm{Q}_k)$. The process noise covariance can be expressed as $ \mathrm{Q}_k=\text{blkdiag}(\mathrm{Q}_k^i, \mathrm{Q}_k^{\boldsymbol{\eta}} )$, where $\mathrm{Q}_k^{\boldsymbol{\eta}}$ represents the modeling uncertainty of the parameter process.

\subsection{Deriving System Noise Covariance}\label{sec:3.2}
 The statistics of the system noise play a vital role in the design and implementation of a Bayesian estimator for a given stochastic system. However, its knowledge is often not readily available to the user \cite{akhlaghi2017adaptive}. The noise associated with the measurement model is predominantly contributed by the sensors that record the measurements, and its statistics are generally provided by the sensor manufacturers. In contrast, process noise represents random variations and unpredictable changes in the system's state that cannot be expressed in the given process dynamics. 

For the underlying process model in the wave-vessel system, we first construct the process uncertainty for the regular wave excitation given in \eqref{reg_model}, and then proceed to derive this for an irregular excitation to express the uncertainty in \eqref{process_dyna}. Notably, this is not a constant-velocity model driven by white noise, as the excitation input is treated as part of the system state. Now, using the model in \eqref{reg_model} and the definitions of $\mathrm{A}^s_k(\boldsymbol{\eta})$ and $\mathbf{B}^s_k(\boldsymbol{\eta})$, we can expand the position state as $x_k = x_{k-1}+\dot{x}_{k-1}T_s$. We can observe that process dynamics do not consider the positional change contributed by instantaneous acceleration during the current sampling time. Thus, the uncertainty in position can be given as \begin{equation}\label{pos_er}
 \Delta x_k = \dfrac{1}{2}{a_{\max}} T_s^2,
 \end{equation}
 where $a_{\max}$ is the maximum acceleration. Similarly, by expanding the velocity state, we have
 \begin{equation*}
 \begin{split}
     \dot{x}_k &= \dot{x}_{k-1} +(p_{k-1}-C\dot{x}_{k-1}-Kx_{k-1})\dfrac{T_s}{M} \triangleq 
      \dot{x}_{k-1} +a_k T_s,
 \end{split}
 \end{equation*}
 where $K$ is the spring constant. It can be observed that the expression for velocity, $\dot{x}_k$, is completely described by system states, and any error in its value can be introduced only by the propagation of uncertainties from other independent variables. In this case, wave excitation could be one such variable, and hence, the uncertainty in velocity can be expressed as
 \vspace{-3mm}
 \begin{equation}\label{vel_er}
     \Delta \dot{x}_k = \lambda \Delta p_{k-1},
 \end{equation}
 \vspace{-6mm}
 
\noindent where $\lambda$ is a scaling factor.
Further, for the third state---the wave excitation---modeled as a regular sinusoid representing the impacting sea wave, its value varies periodically. Moreover, it is expressed recursively in \eqref{input_recur} and does not depend on the other states considered for modeling the system. However, the phase in its expression is unknown and is therefore assigned randomly from the range $[0, \ 2\pi]$. Consequently, it contributes to the uncertainty in the dynamics, whose worst-case value for the $n$th component can be computed as
\begin{equation}\label{input_er}
    \Delta p_k = 2{\tilde{p}_{n}}\sin \left(\frac{\bar{\omega}_nT_s}{2}\right),
\end{equation}
where $\tilde{p}_{n}$ and $\bar{\omega}_n$ are the amplitude and encountered frequency of the $n$th regular component of the wave excitation.  
Assuming a zero-mean Gaussian distribution for the process noise of the regular dynamics in \eqref{reg_model}, the noise covariance for the $n$th component can be computed using \eqref{pos_er}, \eqref{vel_er}, and \eqref{input_er} as
$\mathrm{Q}_k^r({\omega}_n) = \text{diag}\!\big([{\Delta x_k}^2, {\Delta \dot{x}_k}^2, {\Delta p_k}^2]\big).$
Here, the initial values of the quantities are calculated as $\lambda = T_s/M$ and
$\tilde{p}_n({\omega}_n) = Ma_{\max} + C(\omega_n)\dot{x} + x,$
where $M$ and $C$ are given by \eqref{M} and \eqref{C}, with $B = B_0$ and $T = \text{CoG}_z$ as specified in Table~\ref{tab1}. Furthermore, $a_{\max}$, $\dot{x}$, and $x$ are prior values chosen as the maximum magnitudes from a sample of sensor measurements.  
The overall process noise covariance is then written as
$\mathrm{Q}_k = \text{blkdiag}(\mathrm{Q}_k^r({\omega}_1), \ldots, \mathrm{Q}_k^r({\omega}_N), \ \mathrm{Q}^{\boldsymbol{\eta}}),$
where the unknown parameter noise covariance should ideally be zero; however, for numerical stability, it is chosen as
$\mathrm{Q}^{\boldsymbol{\eta}} = \text{diag}([10^{-12}, \ 10^{-13}]).$
Finally, $\mathrm{Q}$ is tuned using the parameters $\lambda$ and $a_{\max}$ to ensure a decreasing normalized innovation square, expressed as $\boldsymbol{\nu}_k^\top {\Sigma_{k|k-1}^{{\mathbf{yy}}^{-1}}}\boldsymbol{\nu}_k$, where $\boldsymbol{\nu}_k = \mathbf{y}_k - \hat{\mathbf{y}}_{k|k-1}$ denotes the innovation, and $\Sigma_{k|k-1}^{\mathbf{yy}}$ is its covariance, computed from \eqref{pred_y} and \eqref{inno_cov}, respectively. The tuning process is stopped once the normalised innovation square falls below a predetermined threshold or exhibits no appreciable change.

\subsection{Input-State-Parameter Estimation}\label{sec:joint estimation}
Having framed the irregular wave-vessel system model and computed the necessary process noise statistics, and assuming the sensor noise information is available, in this section, we formulate the Bayesian estimator for our sea-state estimation problem.  

Our system model in \eqref{irr_process} and \eqref{irr_meas} is linear with respect to the system states and input excitation corresponding to the frequency components of the irregular wave, provided the vessel parameters, $\boldsymbol{\eta} =[B, T]^\top$, are known. However, with the introduction of the unknown parameters $\boldsymbol{\eta}$ in the state vector, the system becomes nonlinear since the other states are expressed as a nonlinear function of the unknown parameters. The resulting problem is an input-state-parameter estimation problem. As discussed in Section~\ref{sec:ckf}, we formulate a CKF to jointly estimate the states along with the wave excitation and parameters for the system represented in \eqref{process_dyna} and \eqref{meas_model}. We describe the two-step, predict and update procedures, for the CKF formulation for the estimation problem next.

The time prediction step of the CKF, as given in Eq.~\eqref{pred_mean1}, can be approximated with the help of a set of cubature points for the system described in \eqref{process_dyna} as follows:
\begin{align}
 \hat{\mathbf{x}}_{k|k-1} & =\dfrac{1}{2n_{\mathbf{x}}}\sum_{j=1}^{2n_{\mathbf{x}}} \mathrm{A}_k(\boldsymbol{\eta}_j)\mathbf{x}_{j,k-1},\label{pred_mean}\\
 \Sigma_{k|k-1} & = \dfrac{1}{2n_{\mathbf{x}}}\sum_{j=1}^{2n_{\mathbf{x}}} (\mathrm{A}_k(\boldsymbol{\eta}_j)\mathbf{x}_{j,k-1}-\hat{\mathbf{x}}_{k|k-1}) (\mathrm{A}_k(\boldsymbol{\eta}_j)\mathbf{x}_{j,k-1}-\hat{\mathbf{x}}_{k|k-1})^\top +\mathrm{Q}_{k-1},\label{pred_cov}\\
 %
   \text{ where} \quad  \mathbf{x}_{j,k-1}& = \hat{\mathbf{x}}_{k-1|k-1} + \xi_j\mathrm{S}_{k-1|k-1}.\label{prior_x}
 \end{align}
 Here, $\xi_j$ is the $j$th cubature point for the given state dimension, $n_{\mathbf{x}}$, and $S_{k-1|k-1}$ is the matrix square-root of covariance $\Sigma_{k-1|k-1}$ such that $\Sigma_{k-1|k-1}= \mathrm{S}_{k-1|k-1}\mathrm{S}_{k-1|k-1}^\top$. 

 The expressions given for posterior mean and covariance in \eqref{post_mean} and \eqref{post_cov}, respectively, are the measurement update. The expressions of posterior mean and covariances 
 can be approximated for our system model in \eqref{process_dyna} and \eqref{meas_model}  with the help of a set of deterministic points, and can be represented as
  \begin{align}
    \hat{\mathbf{y}}_{k|k-1} &= \dfrac{1}{2n_{\mathbf{x}}} \sum_{j=1}^{2n_{\mathbf{x}}} \mathrm{G}_k(\boldsymbol{\eta}_j)\mathbf{x}_{j,k|k-1},\label{pred_y}\\
    \Sigma_{k|k-1}^{\mathbf{yy}} & = \dfrac{1}{2n_{\mathbf{x}}} \sum_{j=1}^{2n_{\mathbf{x}}}(\mathrm{G}_k(\boldsymbol{\eta}_j)\mathbf{x}_{j,k|k-1}-\hat{\mathbf{y}}_{k|k-1})(\mathrm{G}_k(\boldsymbol{\eta}_j)\mathbf{x}_{j,k|k-1}-\hat{\mathbf{y}}_{k|k-1})^\top +R_k,\label{inno_cov}\\
    \Sigma_{k|k-1}^{\mathbf{xy}} & = \dfrac{1}{2n_{\mathbf{x}}} \sum_{j=1}^{2n_{\mathbf{x}}}(\mathbf{x}_{j,k|k-1}-\hat{\mathbf{x}}_{k|k-1})(\mathrm{G}_k(\boldsymbol{\eta}_j)\mathbf{x}_{j,k|k-1}-\hat{\mathbf{y}}_{k|k-1})^\top,\\
   \text{ where} \quad   \mathbf{x}_{j,k|k-1}& = \hat{\mathbf{x}}_{k|k-1} + \xi_jS_{k|k-1}, \quad \text{and} \quad  \Sigma_{k|k-1} =\mathrm{S}_{k|k-1}\mathrm{S}_{k|k-1}^\top \label{predicted_x}
 \end{align}
\vspace{-1mm}
\begin{algorithm}[h!]
\footnotesize
\caption{\textsf{SRCKF} (Joint Estimation of Input-State-Parameters)}\label{algo1}
\textbf{Input:} $\hat{\mathbf{x}}_{k-1|k-1}, \mathrm{S}_{k-1|k-1}, \mathbf{y}_k, \mathrm{A}_k(\boldsymbol{\eta}), \mathrm{G}_k(\boldsymbol{\eta})$\\
\textbf{Output:} $\hat{\mathbf{x}}_{k|k}, \mathrm{S}_{k|k}$\\
\texttt{/* Time update */}
\begin{itemize}[itemsep=0pt,parsep=1pt,topsep=0pt,labelindent=0pt,leftmargin=4mm]
    \item[1.] Compute $\mathbf{x}_{j,k-1}$ and $\hat{\mathbf{x}}_{k|k-1}$ from \eqref{prior_x} and \eqref{pred_mean}, respectively
    \item[2.] Compute $\mathrm{S}_{\mathrm{Q}_{k-1}}$ s.t. $\mathrm{Q}_{k-1}=\mathrm{S}_{\mathrm{Q}_{k-1}}\mathrm{S}_{\mathrm{Q}_{k-1}}^\top$, and $\mathbf{x}_{e,k-1} = \dfrac{1}{\sqrt{2n_{\mathbf{x}}}}\begin{bmatrix}\mathbf{x}_{1,k-1}-\hat{\mathbf{x}}_{k|k-1}& \cdots& \mathbf{x}_{2n_{\mathbf{x}},k-1}-\hat{\mathbf{x}}_{k|k-1}\end{bmatrix}$
    \item[3.] $\mathrm{S}_{k|k-1} = \textsf{Triangularize}\left(\begin{bmatrix} \mathbf{x}_{e,k-1} & \mathrm{S}_{\mathrm{Q}_{k-1}}\end{bmatrix}\right)$
\end{itemize}
\texttt{/* Measurement update */}
\begin{itemize}[itemsep=0pt,parsep=1pt,topsep=0pt,labelindent=0pt,leftmargin=4mm]
    \item[4.] Compute $\mathbf{x}_{j,k|k-1}$ and $\hat{\mathbf{y}}_{k|k-1}$ from \eqref{predicted_x} and \eqref{pred_y}, respectively
    \item[5.] Compute $\mathrm{S}_{\mathrm{R}_{k}}$ s.t. $\mathrm{R}_{k}=\mathrm{S}_{\mathrm{R}_{k}}\mathrm{S}_{\mathrm{R}_{k}}^\top$, and 
    $\mathbf{y}_{e,k|k-1} = \dfrac{1}{\sqrt{2n_{\mathbf{x}}}}\begin{bmatrix}\mathrm{G}_k(\boldsymbol{\eta}_1)\mathbf{x}_{1,k|k-1}-\hat{\mathbf{y}}_{k|k-1}& \cdots& \mathrm{G}_k(\boldsymbol{\eta}_{2n_{\mathbf{x}}})\mathbf{x}_{2n_{\mathbf{x}},k|k-1}-\hat{\mathbf{y}}_{k|k-1}\end{bmatrix}$
    \item[6.]  $\mathrm{S}_{k|k-1}^{\mathbf{yy}} = \textsf{Triangularize}\left(\begin{bmatrix} \mathbf{y}_{e,k|k-1} & \mathrm{S}_{R_{k}}\end{bmatrix}\right)$
    \item[7.] $\mathbf{x}_{e,k|k-1} = \dfrac{1}{\sqrt{2n_{\mathbf{x}}}}\begin{bmatrix}\mathbf{x}_{1,k|k-1}-\hat{\mathbf{x}}_{k|k-1}& \cdots& \mathbf{x}_{2n_{\mathbf{x}},k|k-1}-\hat{\mathbf{x}}_{k|k-1}\end{bmatrix}$;\\
    $\Sigma_{k|k-1}^{\mathbf{xy}} = \mathbf{x}_{e,k|k-1}\mathbf{y}_{e,k|k-1}^\top$
    \item[8.] $\mathrm{L}_k = \left(\Sigma_{k|k-1}^{\mathbf{xy}}/\mathrm{S}_{k|k-1}^{\mathbf{yy}\top}\right)/\mathrm{S}_{k|k-1}^{\mathbf{yy}}$
    \item[9.] Compute $\hat{\mathbf{x}}_{k|k}$ from \eqref{post_mean}, and 
    $\mathrm{S}_{k|k}= \textsf{Triangularize}\left(\begin{bmatrix} \mathbf{x}_{e,k|k-1}-\mathrm{L}_k\mathbf{y}_{e,k|k-1} & \mathrm{L}_k\mathrm{S}_{R_{k}}\end{bmatrix}\right)$
\end{itemize}
\vspace{-1mm}
\end{algorithm}

The complete algorithm for the proposed square-root CKF (SRCKF) is summarised in Algorithm~\ref{algo1}. Notably, we choose the SRCKF to provide numerical stability to the implementation. The \textsf{Triangularize}(.) function represents the process of transforming an augmented matrix into a lower triangular matrix such that the individual square of both matrices results in the same matrix. That is, if $\mathrm{S} =\textsf{Triangularize}(\mathrm{X})$, then, $\mathrm{S}\mathrm{S}^\top = \mathrm{X}\mathrm{X}^\top$, where $\mathrm{S}$ is a lower triangular and square matrix, and $\mathrm{X}$ may be an augmented matrix.

In our formulation, we use parallel mass-spring-damper systems to model the independent single DoF motions of the vessel under the influence of an irregular sea wave. In particular, we consider two systems, one for heave and the other for pitch motions. Although the two systems have different primary states and wave excitations, they share the same vessel and its unknown parameters. In fact, the coefficients of the states dynamics, which are a function of unknown parameters in $\boldsymbol{\eta}$,  are identical for the system. Therefore, it is beneficial to merge the sensors' measurements of the {decoupled} motions to improve the estimates of the shared parameters. Furthermore, we can observe from the process model in~\eqref{process_dyna} that other components of the state and wave excitations are functions of vessel parameters; hence, improving the parameter estimate will result in better estimates for the overall system state. Consequently, we fuse heave and pitch measurements by sharing the estimates of common parameters $\boldsymbol{\eta}$ across two parallel SRCKF algorithms at each time step.

 \begin{algorithm}[h!]
 \footnotesize
     \caption{\textsf{Measurement Fusion}}\label{algo2}
         \textbf{Input:} $\hat{{\tau}}_{k-1|k-1}, \mathrm{S}_{k-1|k-1}^\tau, \mathbf{y}_k^\tau; \hat{\theta}_{k-1|k-1}, \mathrm{S}_{k-1|k-1}^\theta,\mathbf{y}_k^\theta; \mathrm{A}_k(\boldsymbol{\eta}), \mathrm{G}_k(\boldsymbol{\eta})$\\
         \textbf{Output:} $\hat{{\tau}}_{k|k}, \mathrm{S}_{k|k}^\tau; \hat{\theta}_{k|k}, \mathrm{S}_{k|k}^\theta$\\
         \texttt{/*Heave estimation */}
         \begin{itemize}[itemsep=0pt,parsep=1pt,topsep=0pt,labelindent=0pt,leftmargin=4mm]
             \item[1.] $\hat{\boldsymbol{\eta}}_{k-1}^\tau := \hat{\boldsymbol{\eta}}_{k-1}^\theta $
             \item[2.] $\hat{\tau}_{k-1|k-1} :={\begin{bmatrix} \hat{\tau}_{k-1|k-1}^{i^\top} & \hat{\boldsymbol{\eta}}_{k-1}^{\tau^\top}\end{bmatrix}}^\top$
             \item[3.] $\hat{{\tau}}_{k|k}, \mathrm{S}_{k|k}^\tau := \textsf{SRCKF}\left(\hat{{\tau}}_{k-1|k-1}, \mathrm{S}_{k-1|k-1}^\tau, \mathbf{y}_k^\tau, \mathrm{A}_k(\hat{\boldsymbol{\eta}}_{k-1}^\tau), \mathrm{G}_k(\hat{\boldsymbol{\eta}}_{k-1}^\tau)\right)$
         \end{itemize}
         \texttt{/* Pitch estimation */}
         \begin{itemize}[itemsep=0pt,parsep=1pt,topsep=0pt,labelindent=0pt,leftmargin=4mm]
             \item[4.] $\hat{\boldsymbol{\eta}}_{k-1}^\theta := \hat{\boldsymbol{\eta}}_{k}^\tau $
             \item[5.] $\hat{\theta}_{k-1|k-1} :={\begin{bmatrix} \hat{\theta}_{k-1|k-1}^{i^\top} & \hat{\boldsymbol{\eta}}_{k-1}^{\theta^\top}\end{bmatrix}}^\top$
             \item[6.] $\hat{{\theta}}_{k|k}, \mathrm{S}_{k|k}^\theta := \textsf{SRCKF}\left(\hat{{\theta}}_{k-1|k-1}, \mathrm{S}_{k-1|k-1}^\theta, \mathbf{y}_k^\theta, \mathrm{A}_k(\hat{\boldsymbol{\eta}}_{k-1}^\theta), \mathrm{G}_k(\hat{\boldsymbol{\eta}}_{k-1}^\theta)\right)$
         \end{itemize}
\end{algorithm}
\vspace{-3mm}
The idea is to use the posterior estimate of parameters from one system as a prior for the other. To achieve this, we execute our proposed SRCKF algorithm sequentially from heave to pitch and share the parameter estimates during the transition. {We adopt heave-to-pitch fusion rather than pitch-to-heave fusion because, in asynchronous fusion, the more consistent estimates should be used first in the update sequence; heave estimates are generally more consistent, as also reported in~\cite{brodtkorb2023automatic}}.  More specifically, the state vector in \eqref{final_state} is represented as $\tau_k = [\tau_k^{i^\top}, \boldsymbol{\eta}_{k}^{\tau^\top}]^\top$ for heave motion and 
$\theta_k = [\theta_k^{i^\top}, \boldsymbol{\eta}_{k}^{\theta^\top}]^\top$ for pitch motion, where $\boldsymbol{\eta}_{k}^\tau$ and $\boldsymbol{\eta}_{k}^\theta$ denote the parameter estimates obtained from the heave and pitch estimators, respectively, 
at the $k$th time step.  Now, we share the estimated $\hat{\boldsymbol{\eta}}_{k-1}^\theta$ at time step $k-1$ from the pitch estimation algorithm to improve the estimate, $\hat{\boldsymbol{\eta}}_k^\tau$, of the system with heave measurement at time step $k$. Subsequently, $\hat{\boldsymbol{\eta}}_k^\tau$ is used as a prior to estimate $\hat{\boldsymbol{\eta}}_k^\theta$. The proposed fusion algorithm is outlined in Algorithm~\ref{algo2}.

\begin{figure}[h!]
\includegraphics[width=1.0\textwidth]{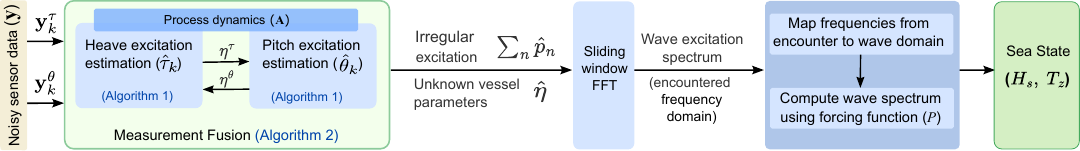}
\caption{An overview of the two-stage approach for joint estimation of vessel parameters and sea states using noisy vessel motion measurements.}
\label{overall-approach}
\end{figure}

We illustrate the overall approach in Figure~\ref{overall-approach}. The irregular wave–vessel system model is used to estimate the irregular wave excitations and unknown vessel parameters using measurements of the vessel's motions. Subsequently, a sliding-window-based FFT is applied to the estimated instantaneous excitations to compute the wave spectrum and associated wave properties $H_s$ and $T_z$.

\section{Deriving A Theoretical Performance Bound}\label{sec:bound}
 The proposed SRCKF for the underlying discrete-time nonlinear system is a suboptimal filter, and it is often difficult to evaluate the achievable performance of a suboptimal filter in a relatively new application where there is no existing benchmark in the literature. To help assess the achievable performance of the proposed algorithm, particularly for unknown irregular wave excitation, here we seek to compute a lower bound for the estimated values.

Among the existing lower bounds on estimation error, as discussed in Section~\ref{sec:performance bounds}, we choose the PCRLB to evaluate the performance of our estimators. The PCRLB, rooted in the Bayesian approach, is particularly suitable for time-varying nonlinear systems where estimators may exhibit some bias in their estimates~\cite{tichavsky1998posterior}. In this section, we first formulate the PCRLB for the system state given in \eqref{final_state} and then use this expression to derive the lower bound for the irregular wave excitation to the wave-vessel system.

 \subsection{Posterior CRLB}
 Following the notation conventions used in this work, in general, the PCRLB for a stochastic system can be determined with the inequality given below \cite{van2004detection}:
 \begin{equation}\label{crlb}
     \Sigma \triangleq \mathbb{E}[(\hat{\mathbf{x}}-\mathbf{x})(\hat{\mathbf{x}}-\mathbf{x})^\top] \geq \mathrm{J}^{-1},
 \end{equation}
  where the Fisher information, $\mathrm{J} \in \Re^{n_{\mathbf{x}}\times n_{\mathbf{x}}}$, is computed as
 \begin{equation}\label{info_mat}
     \mathrm{J}_{ij} = \mathbb{E}\left[-\dfrac{\partial^2 \log P(\mathbf{x},\mathbf{y})}{\partial \mathbf{x}_i \partial \mathbf{x}_j}\right]; \quad i,j = 1, \cdots, n_{\mathbf{x}}.  
 \end{equation}
 Here, $P(\mathbf{x},\mathbf{y})$ is the joint density of the noisy states and measurements. For the underlying discrete-time system, the joint density, $P(\mathbf{x}_{0:k},\mathbf{y}_{1:k})$, can be assumed to be decomposed using the chain rule as
 \begin{equation*}
     \begin{split}
         P(\mathbf{x}_{0:k},\mathbf{y}_{1:k})&= P(\mathbf{x}_k,\mathbf{y}_k,\mathbf{x}_{0:k-1},\mathbf{y}_{1:k-1}),\\
         &=P(\mathbf{y}_k|\mathbf{x}_k,\mathbf{x}_{0:k-1},\mathbf{y}_{1:k-1})P(\mathbf{x}_k|\mathbf{x}_{0:k-1},\mathbf{y}_{1:k-1})P(\mathbf{x}_{0:k-1},\mathbf{y}_{1:k-1}).
     \end{split}
 \end{equation*}
 Assuming that the state $\mathbf{x}_k$ is Markov and depends only on the previous state, also that the current measurement, $\mathbf{y}_k$, conditioned on the current state, $\mathbf{x}_k$, is independent of previous measurements and states, we can write the above joint density as
 \begin{equation}\label{joint__decom}
     P(\mathbf{x}_{0:k},\mathbf{y}_{1:k}) = P(\mathbf{y}_k|\mathbf{x}_k)P(\mathbf{x}_k|\mathbf{x}_{k-1})P(\mathbf{x}_{0:k-1},\mathbf{y}_{1:k-1}).
 \end{equation}
 Aiming to compute the information matrix at the current step using the joint density, the joint state can be split into three parts: current state, previous state, and the rest of the joint state, as below:
 \begin{equation}\label{state_split}
     \mathbf{x}_{0:k} = [\mathbf{x}_{0:k-2}^\top \ \mathbf{x}_{k-1}^\top \ \mathbf{x}_k^\top]^\top
 \end{equation}
 Now, using \eqref{info_mat} and \eqref{state_split}, the information matrix for the joint state $\mathbf{x}_{0:k}$ can be expressed as
 \begin{equation}\label{J}
     \mathrm{J}(\mathbf{x}_{0:k}) = \begin{bmatrix} J_{11} & J_{12} & J_{13}\\ J_{21} & J_{22} & J_{23}\\
     J_{31} & J_{32} & J_{33} \end{bmatrix},
 \end{equation}
 where the block components of the matrix, with the help of \eqref{joint__decom}, are given as \begin{align*}
     J_{11} &= \mathbb{E}\left[-\dfrac{\partial^2 \log P(\mathbf{x}_{0:k-1},\mathbf{y}_{1:k-1})}{\partial \mathbf{x}_{0:k-2} \partial \mathbf{x}_{0:k-2}}\right],\\
     J_{12} & = \mathbb{E}\left[-\dfrac{\partial^2 \log P(\mathbf{x}_{0:k},\mathbf{y}_{1:k})}{\partial \mathbf{x}_{0:k-2} \partial \mathbf{x}_{k-1}}\right] ,\\
     J_{13} & = \mathbb{E}\left[-\dfrac{\partial^2 \log P(\mathbf{x}_{0:k},\mathbf{y}_{1:k})}{\partial \mathbf{x}_{0:k-2} \partial \mathbf{x}_{k}}\right] =0,\\
     J_{21}& = \mathbb{E}\left[-\dfrac{\partial^2 \log P(\mathbf{x}_{0:k},\mathbf{y}_{1:k})}{\partial \mathbf{x}_{k-1} \partial \mathbf{x}_{0:k-2}}\right] = J_{12}^\top,\\
      J_{22} & = \mathbb{E}\left[-\dfrac{\partial^2 \log P(\mathbf{x}_{0:k-1},\mathbf{y}_{1:k-1})}{\partial \mathbf{x}_{k-1} \partial \mathbf{x}_{k-1}}\right] + \mathbb{E}\left[-\dfrac{\partial^2 \log P(\mathbf{x}_{k}|\mathbf{x}_{k-1})}{\partial \mathbf{x}_{k-1} \partial \mathbf{x}_{k-1}}\right],\\
     & = J_{22}^1 + J_{22}^2,\\
       J_{23} & = \mathbb{E}\left[-\dfrac{\partial^2 \log P(\mathbf{x}_{k}|\mathbf{x}_{k-1})}{\partial \mathbf{x}_{k-1} \partial \mathbf{x}_{k}}\right], 
     \end{align*}
     \begin{equation}\label{compute_j}
     \begin{split}
     J_{31} & = \mathbb{E}\left[-\dfrac{\partial^2 \log P(\mathbf{x}_{0:k},\mathbf{y}_{1:k})}{\partial \mathbf{x}_{k} \partial \mathbf{x}_{0:k-2}}\right] =0,\\
     J_{32} & = \mathbb{E}\left[-\dfrac{\partial^2 \log P(\mathbf{x}_{k}|\mathbf{x}_{k-1})}{\partial \mathbf{x}_{k} \partial \mathbf{x}_{k-1}}\right] = J_{23}^\top,\\
     J_{33} & = \mathbb{E}\left[-\dfrac{\partial^2 \log P(\mathbf{x}_{k}|\mathbf{x}_{k-1})}{\partial \mathbf{x}_{k} \partial \mathbf{x}_{k}}\right] + \mathbb{E}\left[-\dfrac{\partial^2 \log P(\mathbf{y}_{k}|\mathbf{x}_{k})}{\partial \mathbf{x}_{k} \partial \mathbf{x}_{k}}\right].
     \end{split}
 \end{equation}
 Clearly, as the time step $k$ grows, the computation of the information matrix for the joint state is computationally infeasible, and this necessitates expressing it recursively. If the inverse of the information matrix in \eqref{J}, which is the lower bound of squared error for the joint state, can be expressed as 
 \begin{equation*}\label{j-}
 {\mathrm{J}(\mathbf{x}_{0:k})}^{-1} = \begin{bmatrix} I_{11} & I_{12} & I_{13}\\ I_{21} & I_{22} & I_{23}\\
     I_{31} & I_{32} & I_{33} \end{bmatrix},
 \end{equation*}
 then, using the matrix inverse expansion \cite{lu2002inverses, tichavsky1998posterior}, $I_{33}$, which is the lower bound of squared error for state $\mathbf{x}_k$, can be written as
 \begin{equation*}
 \begin{split}
     I_{33} &= \left(J_{33} -[J_{13} \ J_{23}] {\begin{bmatrix} J_{11}& J_{12}\\ J_{21} & J_{22}\end{bmatrix}}^{-1} [J_{13} \ J_{23}]^\top\right)^{-1}\\
     & = (J_{33} - J_{23}[J_{22}^1 +J_{22}^2-J_{21}J_{11}^{-1}J_{12}]^{-1}J_{32})^{-1}.
     \end{split}
 \end{equation*}
 If ${\mathrm{J}(\mathbf{x}_{0:k})}^{-1}$ is the lower bound of the joint state $\mathbf{x}_{0:k}$, then $I_{33}$ is the lower bound of the component $\mathbf{x}_k$, we can write the inverse of $I_{33}$ as the information submatrix for state $\mathbf{x}_k$, and it is expressed below:
 \begin{equation}\label{J_k}
     J_k = J_{33} - J_{23}[J_{22}^1 +J_{22}^2-J_{21}J_{11}^{-1}J_{12}]^{-1}J_{32}.
 \end{equation}
 Similarly, computing the information submatrix $\mathrm{J}_{k-1}$ for the state $\mathbf{x}_{k-1}$ while writing $\mathbf{x}_{0:k-1} = [\mathbf{x}_{0:k-2}^\top \ \mathbf{x}_{k-1}^\top]^\top$, we have 
 \begin{equation}\label{J_k-1}
     \mathrm{J}_{k-1} = J_{22}^1 -J_{21}J_{11}^{-1}J_{12}.
 \end{equation}
 Substituting \eqref{J_k-1} into \eqref{J_k} yields
 \begin{equation}\label{rec_j}
     \mathrm{J}_k =  J_{33} - J_{23}[J_{k-1} +J_{22}^2]^{-1}J_{32}.
 \end{equation}
 
 \subsection{PCRLB for Estimated System Inputs}
 Now, for the underlying Gaussian system represented in \eqref{process_dyna} and \eqref{meas_model}, the logarithmic transitional and likelihood densities are given as 
 \begin{equation}\label{log_dens}
 \begin{split}
     -\log P(\mathbf{x}_k|\mathbf{x}_{k-1}) &= (\mathbf{x}_k-\mathrm{A}_k(\boldsymbol{\eta})\mathbf{x}_{k-1})^\top \mathrm{Q}_{k-1}^{-1}(\mathbf{x}_k-\mathrm{A}_k(\boldsymbol{\eta})\mathbf{x}_{k-1}),\\
     -\log P(\mathbf{y}_k|\mathbf{x}_k) &= (\mathbf{y}_k-\mathrm{G}_k(\boldsymbol{\eta})\mathbf{x}_k)^\top \mathrm{R}_k^{-1} (\mathbf{y}_k-\mathrm{G}_k(\boldsymbol{\eta})\mathbf{x}_k).
     \end{split}
 \end{equation}
Then, using \eqref{compute_j}, \eqref{rec_j} and \eqref{log_dens}, the information matrix for the underlying system can be recursively computed at each time step starting with $J_0 = \Sigma_0^{-1}$. The inverse of $J_k$ will provide a lower bound on the expectation of the squared error of the estimate $\hat{\mathbf{x}}_{k|k}$. In particular, the state $\mathbf{x}_k$ contains unknown vessel parameters, $\boldsymbol{\eta}$, and regular wave excitations, $p_{1,k}, \cdots, p_{N,k}$; however, our final objective is to provide a lower bound on irregular wave excitation, which is the sum of all regular wave excitations, that is, $p_k^i = \sum_{n=1}^N p_{n,k}$.

 The inverse of the information matrix, as evident in \eqref{crlb}, can be represented as the expectation of squared error. Adhering to this assumption, we consider the value of irregular wave excitation, $p^{i\ast}_k$, which corresponds to its lower bound and can be expressed as follows:
 \begin{equation*}
     {\mathrm{J}(p_k^i)}^{-1} = \mathbb{E}[(p_k^i-p_k^{i\ast})(p_k^i-p_k^{i\ast})^\top] \leq \mathbb{E}[(p_k^i-\hat{p}_k^{i})(p_k^i-\hat{p}_k^{i})^\top],
 \end{equation*}
 where $\hat{p}_k^i$ is the estimated value of irregular wave excitation. Expressing the above inverse of the information matrix in terms of regular wave excitations, we can write
 \begin{equation}\label{irr_j1}
     \begin{split}
         {\mathrm{J}(p_k^i)}^{-1} &= \mathbb{E}\left[\left(\sum_{n=1}^Np_{n,k}-\sum_{n=1}^Np_{n,k}^\ast\right)\left(\sum_{n=1}^Np_{n,k}-\sum_{n=1}^Np_{n,k}^\ast\right)^\top\right],\\
         &= \mathbb{E}\left[\left(\sum_{n=1}^N(p_{n,k}-p_{n,k}^\ast)\right)\left(\sum_{m=1}^N(p_{m,k}-p_{m,k}^\ast)^\top\right)\right],
     \end{split}
 \end{equation}
 where $p_{n,k}^\ast$ is the value of the $n$th regular wave excitation corresponding to the lower bound, ${\mathrm{J}(p_{n,k})}^{-1}$, which is already computed as part of ${\mathrm{J}_k}^{-1}$. Then, defining ${\mathrm{J}(p_{n,m,k})}^{-1}=\mathbb{E}[(p_{n,k}-p_{n,k}^\ast)(p_{m,k}-p_{m,k}^\ast)^\top]; \ n, m = 1, \cdots, N$, where ${\mathrm{J}(p_{n,m,k})}^{-1}$ is the $(3n,3m)$th element of the matrix ${\mathrm{J}_k}^{-1}$, the lower bound of irregular wave excitation in~\eqref{irr_j1} can be rewritten as
 \begin{equation}\label{irr_j}
     {\mathrm{J}(p_k^i)}^{-1} = \sum_{n=1}^N \sum_{m=1}^N {\mathrm{J}(p_{n,m,k})}^{-1}.
 \end{equation}
The lower bound computed on the irregular wave excitation,  ${\mathrm{J}(p_k^i)}^{-1}$, can be considered as a reference to evaluate the performance of the implemented estimators.

 \section{Experiments and Results}\label{sec:exp}
In this section, we describe the series of extensive experiments conducted to validate the modelling approach and the SRCKF-based joint estimation formulation proposed in Section~\ref{srckf} for the system model we developed in Section~\ref{sys_mod}. To validate our approach, {the sea vessel described in Section~\ref{VesselSpec} is used to generate measurements in different sea conditions for the two sets of measurement data:}  
 \begin{itemize}[itemsep=1pt,parsep=1pt,topsep=1pt,labelindent=0pt,leftmargin=5mm] 

    \item Synthetic Data---We generate measurement data for vessel responses to test the effectiveness of our proposed estimation and fusion algorithms. Importantly, synthetic data provides a reliable means to validate 
    a method across a range of settings, where the ground truth is reliably and easily obtained. 
    \item High-Fidelity Sea Vessel Data---To understand the performance of the proposed formulation in real-world operational settings, we consider a dataset generated from a high-fidelity simulation of a sea vessel under the impact of an irregular sea to obtain measurement data expected from on-vessel sensors. Importantly, the high-fidelity simulation not only allows for recreating a real-world vessel's response to irregular seas but also enables the acquisition of accurate ground-truth data for validation. 
  \end{itemize}
To compare the performance of our proposed joint estimation algorithm, we also implement the existing linear formulation equivalent to the current state-of-the-art methods~\cite{pascoal2017estimation, micaela}. We denote this baseline as \textit{KF (known parameters)}. The linear formulation assumes complete knowledge of the vessel's parameters and uses a Kalman filter formulation to estimate the irregular wave properties.

\subsection{Performance Evaluation}\label{sec: perform eval}
To evaluate the effectiveness of the proposed techniques, we compute and compare the critical sea parameters: i)~the significant wave height $H_s$; and ii)~the zero up-crossing period $T_z$ from the estimated spectra with their true values used to generate irregular sea waves impacting the vessel. Notably, in the case of high-fidelity data, we have ground truths for sea parameters only. On the other hand, we are generating the ground truth data for wave excitations for the synthetic measurement data.  
Hence, using the synthetic dataset, we further: i)~compare the root mean square errors (RMSE) of the wave excitation from the two approaches to evaluate the effectiveness of the proposed estimator; and ii)~compute the PCRLB of wave excitation and plot it along with the RMSE values for the synthetic data in order to evaluate the performance of our estimator against the theoretical lower bound.

 Once the instantaneous irregular wave excitation is estimated by using the proposed formulation, we used the FFT method with a sliding window algorithm~\cite{henry2023ultra, richardson2018sliding, jing2024time} to compute the spectral distribution of the regular wave excitations in the encountered frequency domain. Here, the window length for the FFT is calculated as $L_w = {2\pi}/{\min(\omega_{\min}, \Delta \omega_{\min})} $, where $\Delta \omega_{\min}$ and $\omega_{\min}$ denote the minimum frequency interval and the minimum frequency used in the estimation. Further, using the relation in \eqref{doppler}, encountered wave frequencies are mapped into wave frequencies, and with the relationships expressed for the forcing functions in  \eqref{force_w}, and \eqref{force_theta}, the wave excitation amplitudes are mapped back into the corresponding wave amplitudes. Subsequently, the sea wave spectrum $S(\omega_n)$ is computed as $\tilde{\zeta_n}^2/(2\Delta \omega_n)$.

The sea parameters are derived from the spectral moments and the spectral peak using the following equations~\cite{jensen2001load}: $H_s = 4\sqrt{m_0}$, {and the zero up-crossing period is estimated using two standard approaches: (i) $T_z$-I $= T_p/1.41$, and (ii) $T_z$-II $= 2\pi \sqrt{m_0/m_2}$. Here, $m_0$ and $m_2$ denote the zeroth and second-order spectral moments, respectively, while $T_p$ represents the peak period corresponding to the dominant wave component.}  

\subsection{Evaluations with Synthetic Data}
We use a consistently-employed sensor measurement generation method for a sea vessel in the field~\cite{kim2019real,kim2023real,andreas}. Using known system parameters and wave spectrum density, we simulate onboard sensor measurements (vessel response) for irregular sea waves. We detail data generation, experimental settings and the results from our evaluation obtained from a series of Monte-Carlo simulation experiments next.  

\subsubsection{Wave Spectrum}
For synthetic data generation, we use a Bretschneider spectrum~\cite{bretschneider1959wave} to represent the energy distribution of an irregular wave impacting the vessel. Bretschneider spectrum is a two-parameter density used to depict a {fully developed sea} with varying energy levels by appropriately selecting different combinations of its parameters, significant wave height ($H_s$) and {zero up-crossing period} ($T_z$). The  spectrum is described by \cite{micaela, bretschneider1959wave}
 \begin{equation}
     S_{B}(\omega) = \dfrac{a}{\omega^5}\exp \left(\dfrac{-b}{\omega^4}\right),
 \end{equation}
 where $a = \dfrac{H_s^2}{4\pi}\left(\dfrac{2\pi}{T_z}\right)^4$ and $b=\dfrac{1}{\pi}\left(\dfrac{2\pi}{T_z}\right)^4$. Using a Bretschneider spectrum to generate wave excitations, we consider two unidirectional spectral forms to demonstrate the effectiveness of our formulation.

\begin{wraptable}{r}{0.55\columnwidth}
\vspace{-10mm}
\caption{Vessel specifications.}\label{tab1}
\vspace{-5mm}
\begin{center}
\resizebox{0.55\columnwidth}{!}{
    \begin{tabular}{| >{\centering\arraybackslash}m{6.4cm} | >{\centering\arraybackslash}m{1.6cm} |>{\centering\arraybackslash}m{1.6cm}|}
        \hline
        \textbf{Length (m)} & $L$&7.00 \\
        \hline
        \textbf{Breadth (m)} & $B_0$& 2.77\\
\hline
        \textbf{Draught (m)} & $T$ & 0.35 \\
        \hline
            \textbf{ Longitudinal Center of Gravity (m)} & $\mathrm{CoG}_x$ & 2.11 \\
            \hline
             \textbf{Vertical Center of Gravity   (m)} & $\mathrm{CoG}_z$ & 0.79 \\
             \hline
    \end{tabular}
    }
\end{center}
\vspace{-2mm}
\end{wraptable}

 \subsubsection{Vessel Specifications}\label{VesselSpec}
We used specifications of a vessel from an experimental data collection study conducted at the Australian Maritime College, University of Tasmania's (AMC-UTAS) Towing Tank facility. Considering the specification of the sea vessel used in the study, we can maintain consistency with the evaluation in the high-fidelity dataset. The simulator for the high-fidelity dataset was validated against the sea vessel in wave tank experiments at AMC-UTAS, as discussed in Section~\ref{sec:pan-model-validation}. 

 \begin{figure}[ht!]
	\centering
    \begin{subfigure}{.55\columnwidth}
	\includegraphics[width=\columnwidth]{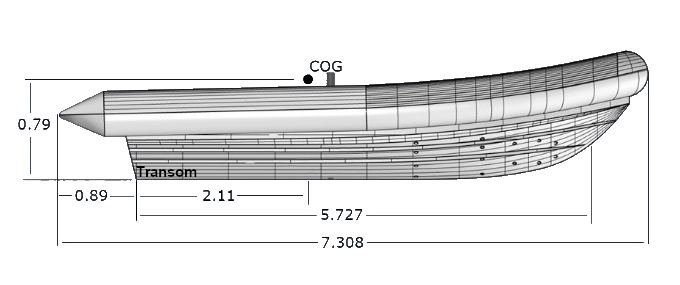}
	\caption{}
	\label{fig4}
    \end{subfigure}
    \begin{subfigure}{.43\columnwidth}
        \includegraphics[width=\columnwidth,height=1.3in,keepaspectratio]{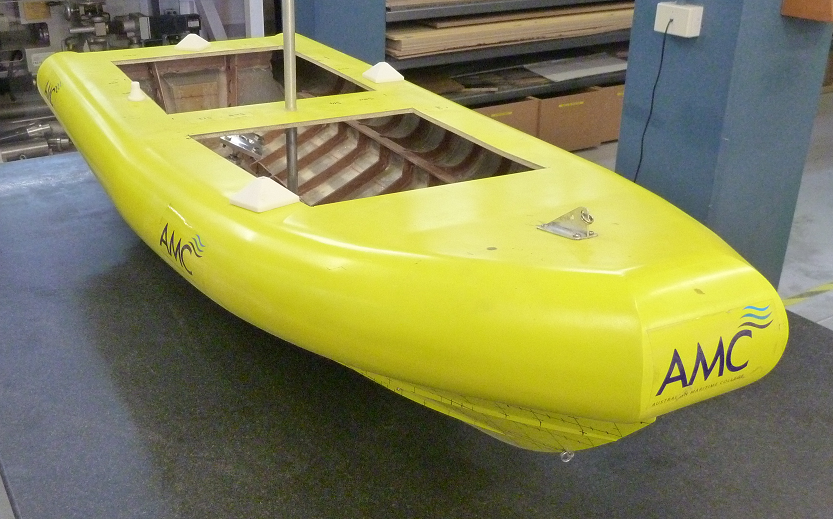}
    \caption{}
    \label{ModelImage }
    \end{subfigure}
    \vspace{-3mm}
    \caption{(a). Vessel specifications expressed in meters. (b). The scaled vessel model (1:5)}
\end{figure}

\subsubsection{Measurement Generation}
To generate the vessel response data using the true states of the wave-vessel system, we assume that all vessel parameters, including $\boldsymbol{\eta}$, are known. More specifically, we execute the expressions given in \eqref{irr_process} and \eqref{irr_meas} recursively after starting with an initial value, $\mathbf{x}_0^i = [\mathbf{x}_{1,0}^{r\top}, \cdots, \mathbf{x}_{N,0}^{r\top}]^\top,$ where $\mathbf{x}_{n,0}^r=[\mathbf{0}^\top, \ p_0]$. Here, $\mathbf{x}_{n,0}^r$ is $n$th frequency component of the irregular state $\mathbf{x}_0^i$, and $p_0$ is the wave excitation input to the heave or pitch system, computed at time step $k=0$. Further, in order to construct the time-dependent wave excitation signal for both systems, we need to consider its three parameters, frequency ($\omega$), phase ($\phi$), and amplitude ($\tilde{\zeta}P$). The frequencies are considered in the range of $[0.20 \ 1.60]$ rad/s with a total number of components $N=30$; the wave amplitudes are calculated using the expression, $\tilde{\zeta}_n = \sqrt{2\Delta \omega_nS_B(\omega_n)}$ for $n$th frequency component; the expressions of forcing functions ($P_n$) contributed due to the geometry of vessel are given in \eqref{force_w} and \eqref{force_theta}, and they can be computed for every frequency $\omega_n$; the phase difference among the components are assigned randomly following a uniform distribution in the range $0$ to $2\pi$ radians. Then, the displacement, velocity, and acceleration measurements for the vessel’s heave and pitch motions under irregular seas are generated with added zero-mean Gaussian noise. The standard deviation in generated measurements  is considered as  $([1.23\mathrm{m},\ 1.33\mathrm{m/s},\ 2.89\mathrm{m/s^2}]\times10^{-2})$ for heave estimation, and $([3\mathrm{rad},\ 1.5\mathrm{rad/s},\ 2.89\mathrm{rad/s^2}]\times10^{-3})$ for pitch estimation, respectively.
 
 \subsubsection{Settings}
 In our experiments, all filters---KF (known parameters) and SRCKF---using heave and pitch measurements, are initialised with $\hat{\mathbf{x}}_0 =  [\hat{\mathbf{x}}_0^{i^\top}, \hat{\boldsymbol{\eta}}_0^\top]^\top$, where the initial estimates and covariances of $n$th regular components of $\hat{\mathbf{x}}_0^i$ are $[0,0,0]$ and $\text{diag}([10^2, 10^2, 10^2]); \forall n =1,\cdots, N$. In evaluating our proposed method with the SRCKF, there is no direct information available about the unknown parameter $\boldsymbol{\eta}$ in the measurements. Consequently, for a given set, an infinite number of combinations of $B$ and $T$ is possible. To address this, we assume prior distributions for these two parameters---a reasonable approach in real-world settings where their actual values may be uncertain or vary, but their approximate ranges are typically known. Specifically, we consider a uniform distribution, $\hat{\boldsymbol{\eta}}_0 \sim [\mathcal{U}([B_0/2, \ 2B_0/3])^\top, \mathcal{U}([\mathrm{CoG}_z/8, \ \mathrm{CoG}_z])^\top]^\top$. For the KF (known parameters), we need only $\hat{\mathbf{x}}_0^i$ and its corresponding covariance. The process noise covariance for SRCKF, $\mathrm{Q}_k$ is constructed using the noise covariances of each regular component together with $\mathrm{Q}^{\boldsymbol{\eta}}$ as explained in Section~\ref{sec:3.2}. In contrast, for the KF (known parameters), the process noise covariance includes only $\mathrm{Q}_k^i$.
 Furthermore, for the estimation problem, we consider a frequency range of $0.40$~rad/s to $1.50$~rad/s to reduce computational effort by lowering the overall dimensionality of the system. This truncation at both ends of the range improves numerical stability, as the spectrum values are low at the extremes; moreover, it also reflects the fact that the estimator does not have access to the full frequency components.

The true wave energy distribution, employed for synthetic data generation, adhering to the Bretschneider spectrum, is constructed with two sets of vessel headings to generate two scenarios for collecting measurement data for a sea vessel to demonstrate generality:
 \begin{itemize}[itemsep=1pt,parsep=1pt,topsep=1pt,labelindent=0pt,leftmargin=5mm]
     \item $H_s = 1.25$~m and $T_z=7$~s {when encountering head seas ~($\beta =\pi$)}; 
     \item $H_s = 1.25$~m and $T_z=7$~s {when encountering beam seas ~($\beta=\pi/2$ radians).}  
 \end{itemize}
 \begin{wrapfigure}{r}{0.44\columnwidth}
    \centering
    \includegraphics[width=0.44\columnwidth]{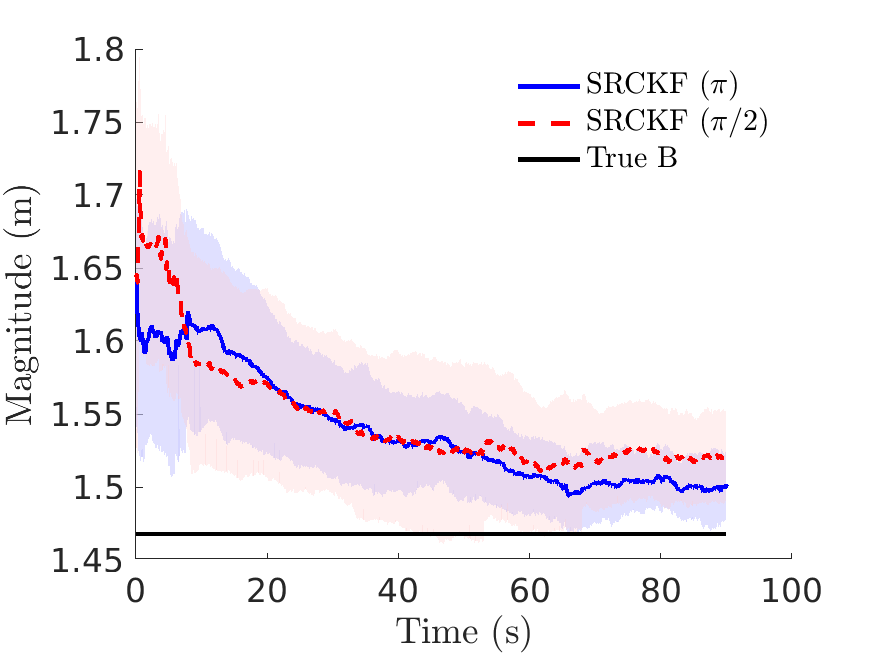}%
\caption{{Estimation of vessel parameter breadth $B$ with $\pm1\sigma$ using the proposed joint estimation approach.}}
    \label{case1_b}%
\vspace{-7mm}
\end{wrapfigure}

The measurement data are recorded at a sampling rate of $25$~Hz, and we consider the vessel moving with a forward speed of $V=4$~m/s under two different relative headings. Notably, the two relative headings of $\pi$ and $\pi/2$ are selected primarily to create two conditions where both pitch and heave measurements are available for estimation, and when only heave measurements are available for estimation---there is no pitch response from the vessel if the relative heading is $\pi/2$.
In the following, we report and discuss the results for the two scenarios.

\begin{wrapfigure}{r}{0.44\columnwidth}
\vspace{-11mm}
    \includegraphics[width=0.44\columnwidth]{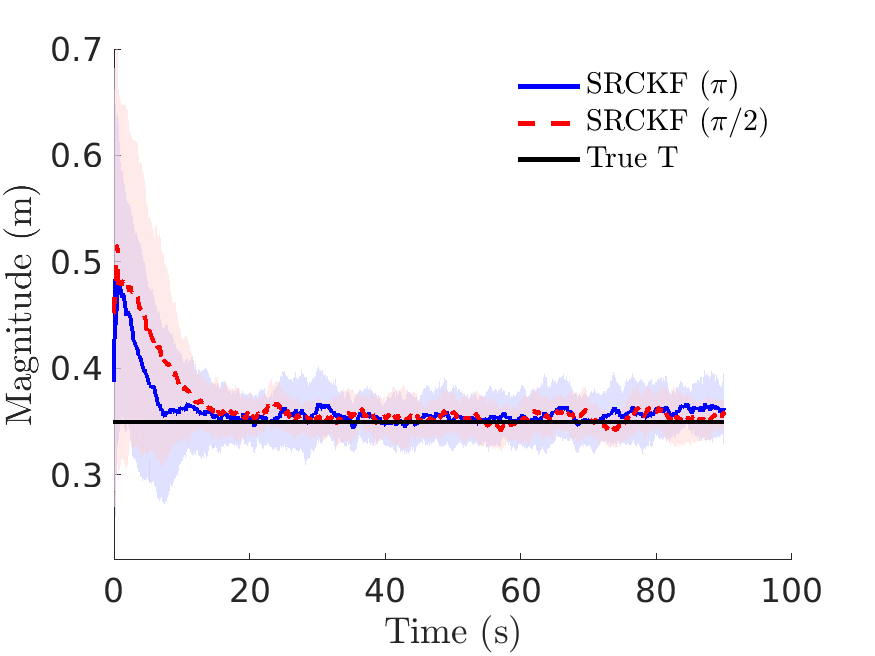}%
\caption{{Estimation of vessel parameters draught $T$ with $\pm1\sigma$ using the proposed joint estimation approach.}}
    \label{case1_t}%
    \vspace{-5mm}
\end{wrapfigure}%
\subsubsection{Results and Discussion}\label{sec:case-i-results} 
We analyse the simulation results for over $50$ Monte Carlo (MC) runs, each with an estimation duration of $90$s, to provide a reliable mean and variance, as $\pm1\sigma$, for filter estimates to understand the consistency and reliability of the proposed estimator. 
The unknown vessel parameters $B$ and $T$ are estimated and plotted for { the head ($\pi$) and beam ($\pi/2$) seas} in Figures~\ref{case1_b} and \ref{case1_t}. As it can be observed in Equations \eqref{M} and \eqref{C}, {$T$ is directly proportional to $M(\boldsymbol{\eta})$ and is a component in $C(\boldsymbol{\eta})$ coefficient}; hence, it is strongly reflected in the vessel response through both the acceleration and velocity components in the measurement vector. 
In contrast, $B$ is weakly related to only the pseudo damping constant and is only reflected in the vessel response through the velocity component in the measurement.  
These relationships of the parameters with sensor measurements are clearly reflected in plots, where $T$ converges quickly around its true value, while $B$ converges relatively slower.  
\begin{figure}[h!]
\centering
\includegraphics[width=\textwidth]{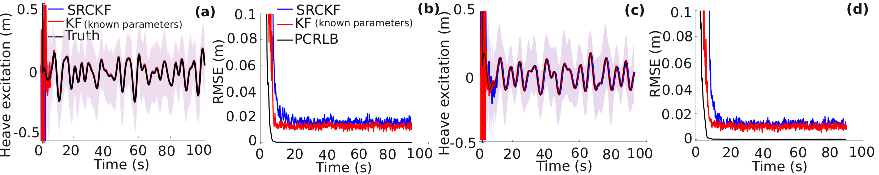}
\caption{{Estimation of: (a) heave excitation with $\pm1\sigma$ uncertainty and (b) its RMSE and PCRLB under head seas and (c) heave excitation with $\pm1\sigma$ uncertainty and (d) its RMSE and PCRLB under beam seas}}.
\label{case1_heave}
\vspace{-5mm}
\end{figure}
\begin{figure}[h!]
\centering
\includegraphics[width=\textwidth]{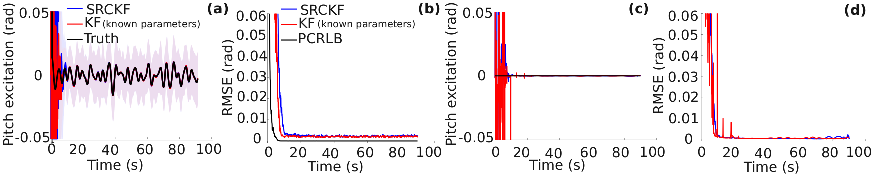}
\caption{{Estimation of: (a) pitch excitation with $\pm1\sigma$ uncertainty and (b) its RMSE and PCRLB under head seas and (c) pitch excitation with (d) its RMSE under beam seas. Notably, there is no truth for pitch excitation under beam seas, and PCRLB cannot be computed.}}
\label{case1_pitch}
\end{figure}  

{Figure~\ref{case1_heave} presents the reconstructed heave excitations, the RMSE of their estimates, and their performance relative to the PCRLB for the two headings.} As expected, KF (known parameters) converges faster around its truth, whereas the proposed SRCKF is slightly slower, owing to the convergence time of unknown parameters. 
Furthermore, the small difference between the RMSE of the two filters is attributed to the unknown vessel parameters, while the gap between the RMSE of the two filters and the PCRLB represents the partial observability of the underlying system. Recall, the unobserved state vector contains the regular excitations with different wave harmonics and random phase differences among them, while the measurement vector represents the irregular vessel responses. Then, the estimation of the individual regular components of the state vector does not follow their true values; instead, the filters correct their sum towards its true value using the corresponding irregular measurement. However, the SRCKF formulation, which jointly estimates the vessel parameters and sea state, achieves an RMSE with a similar margin to that achieved by the KF formulation with known parameters when compared to the PCRLB.

 Figure~\ref{case1_pitch} reports the pitch excitation estimates and their RMSE, along with the truth and PCRLB for the two encountered scenarios. After initial fluctuations, both filters converge around the truth in the case of pitch reconstructions, and their RMSE, on the right side, reflect these observations.
Notably, the PCRLB for pitch estimation cannot be computed for the heading of $\pi/2$ as there is no truth when waves impact the vessel perpendicular to its heading. It is also interesting to note in the case with heading $\pi/2$ when there is only noise in the pitch measurements, KF (known parameters) reacts sharply in the early stage of estimation (when the covariance is still relatively large) to some outlier noise sequences, in contrast, the SRCKF, which has a slower response because of its nonlinear measurement model is able to handle the noise better.

\begin{figure}[h!]%
	\centering
	\begin{subfigure}{.44\columnwidth}
		\includegraphics[width=\columnwidth]{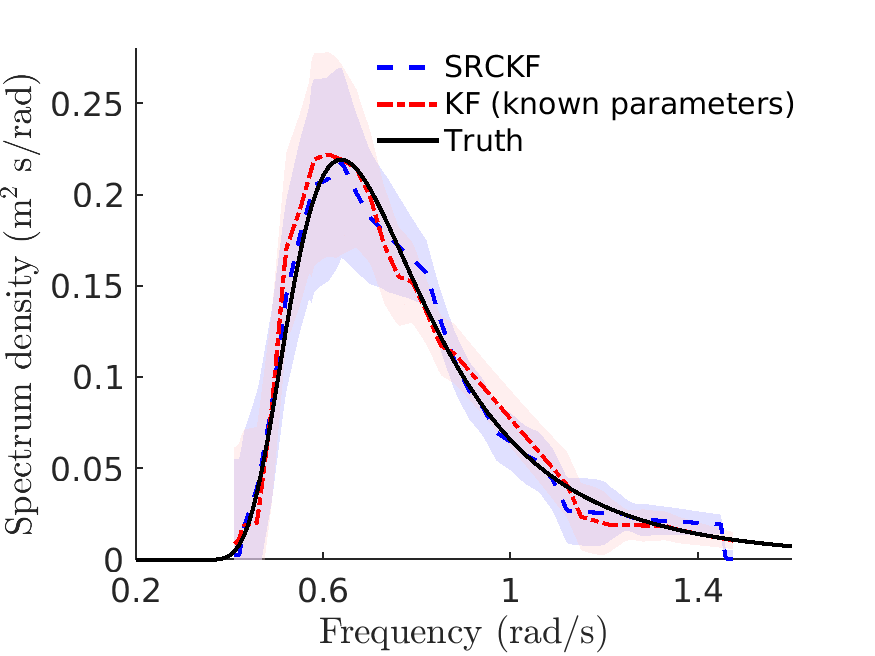}%
		\caption{{Head sea}}%
		\label{case1_spec_180}%
	\end{subfigure}\hfill%
	\begin{subfigure}{.44\columnwidth}
		\includegraphics[width=\columnwidth]{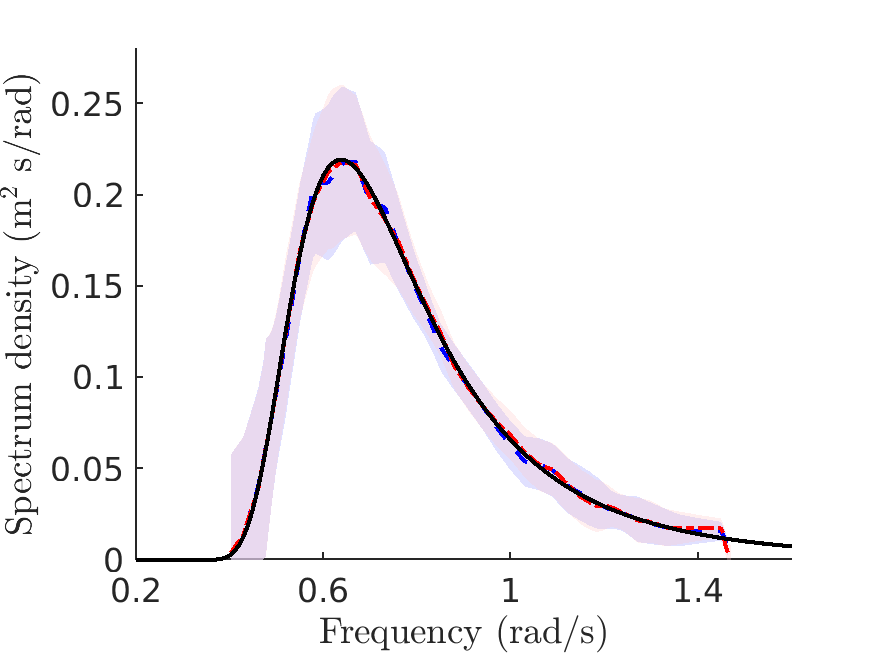}%
		\caption{{Beam seas.}}%
		\label{case1_spec_90}%
	\end{subfigure}%
    \vspace{-2mm}
	\caption{{Estimation of wave spectra under  head seas and beam seas (different relative heading directions).}}
	\label{case1_spec}
\end{figure}

The wave spectra computed, as discussed in Section~\ref{sec: perform eval}, for the two different headings are plotted in Figure~\ref{case1_spec} along with the corresponding true spectra.
Although spectrum estimation uses only the heave response in Figure~\ref{case1_spec_90}, it is interesting to note that the computed values appear to be consistently closer to their truth than in the case where both heave and pitch measurements contribute to the estimation problem. 
This is mainly attributed to the stretching of the wave frequency range---see \eqref{doppler}---in its encounter domain when the wave direction (heading) is $\pi$ (head seas); this does not occur in the case of heading $\pi/2$ (beam seas), and the wave components remain the same in both domains. 

\begin{wraptable}{l}{0.6\columnwidth}
\vspace{-3mm}
\centering
\caption{Comparing estimated results versus the truth for the spectrum and vessel parameters under the two relative headings. Here, KF (known) parameters method assumes knowledge of vessel parameters ($B$ and $T$).}
\label{tab2}
\resizebox{0.6\columnwidth}{!}{
\begin{tabular}{|c|c|c|c|c|c|c|} 
\cline{2-7}
\multicolumn{1}{c|}{}                           & \multirow{2}{*}{\textbf{Parameters} } & \multirow{2}{*}{\textbf{Truth} } & \multicolumn{2}{c|}{\textbf{SRCKF} }                    & \multicolumn{2}{c|}{\textbf{KF (known)} }              \\ 
\hhline{~~~----|}
\multicolumn{1}{c|}{}                           &                                       &                                  & {\cellcolor[rgb]{0.894,0.894,0.894}}$\pi$    & $\pi/2$  & {\cellcolor[rgb]{0.894,0.894,0.894}}$\pi$  & $\pi/2$   \\ 
\hhline{|=======|}
\multirow{2}{*}{\textbf{Vessel Parameters} }    & $B$ (m)                               & $1.47$                           & {\cellcolor[rgb]{0.894,0.894,0.894}}$1.51$   & 1.52     & {\cellcolor[rgb]{0.894,0.894,0.894}} -      & -         \\ 
\hhline{|~------|}
                                                & $T$ (m)                               & $0.35$                           & {\cellcolor[rgb]{0.894,0.894,0.894}}$0.35 $  & 0.35     & {\cellcolor[rgb]{0.894,0.894,0.894}}-      & -         \\ 
\hline
\multirow{3}{*}{ \textbf{Sea Wave Parameters} } & $H_s$ (m)                             & $1.25$                           & {\cellcolor[rgb]{0.894,0.894,0.894}}$1.23$   & $1.22$   & {\cellcolor[rgb]{0.894,0.894,0.894}}1.24   & 1.23      \\ 
\hhline{|~------|}
                                                & $T_z$-I (s)                           & $7.0$                            & {\cellcolor[rgb]{0.894,0.894,0.894}}$7.2$    & $6.9$    & {\cellcolor[rgb]{0.894,0.894,0.894}}$7.3$  & $6.9$     \\ 
\hhline{|~------|}
                                                &{ $T_z$-II (s)}                          & {$7.0$}                            & {\cellcolor[rgb]{0.894,0.894,0.894}}{$7.8$}    & {$7.8$ }   & {\cellcolor[rgb]{0.894,0.894,0.894}}{$7.9$ } & {$7.8$}     \\
\hline
\end{tabular}
}
\end{wraptable}

Using the estimated spectra from the proposed joint estimation method and the method with known parameters, we calculated the sea wave parameters: i)~significant wave height ($H_s$); and ii)~zero up-crossing time ($T_z$), as shown in Table~\ref{tab2}. Importantly, both methods---the joint estimation with SRCKF and the KF with known vessel parameters---performed similarly in estimating the sea wave parameters compared to the ground truth. The results show an accuracy better than $2.5\%$ for $H_s$, $4\%$ for $T_z$-I, and $13\%$ for $T_z$-II for both filters, suggesting that jointly estimating the unknown parameters does not significantly compromise the accuracy of spectrum estimation. This confirms the reliability of the proposed method in approximating sea wave parameters, even when some vessel parameters are not entirely known. 

\subsection{High-Fidelity Dataset Generation and Validation}\label{sec:high-fed-data-gen}
The high-fidelity data sets used to test the algorithms were generated using the numerical modeling tool, PanShipL. PanShipL is a semi-linear time domain panel method for predicting hydrodynamic loads and seakeeping behaviour of high-speed craft operating in waves. The tool has been developed by the Maritime Research Institute Netherlands (MARIN) under the FAST3 joint industry project. This was an international collaborative consortium comprised of Damen Shipyards, Defence Material Organisation (Netherlands), Delft University of Technology, MARIN and the Defence Science and Technology Group.

In this section, we describe the systematic method used to validate the PanShipL simulation model employed for generating high-fidelity data for a sea-vessel under various sea conditions. The validated numerical model within PanShipL can subsequently be used to explore a wider range of sea conditions and vessel velocities and scale up to a full sea wave spectrum, such as a Bretschneider spectrum. This approach significantly reduces the challenges associated with real-world testing, while providing the flexibility to generate data under various scenarios for both vessels and sea states.
Subsequently, we evaluate our approach using the generated datasets.

\subsubsection{Model Testing Data Generation}
A model test program was undertaken at the AMC-UTAS' Towing Tank using a 1:5 scale model of a 7~m waterjet propelled RHIB, as previously detailed in Section~\ref{VesselSpec}. Fibreglass was used to construct the model shown in Figure~\ref{ModelImage }, which included the chines and lifting strakes, but did not include the waterjet or intake structure. The conditions tested in the program comprised of equivalent full scale calm water velocities of between 2 and 20 knots, in 2 knot increments, and regular sea velocities of 6, 8 and 10 knots, with wave heights and frequencies, as listed in Table \ref{modelTestConditions}. 

\begin{wraptable}{l}{0.6\columnwidth}
\centering
\caption{Scale model test conditions.}
\vspace{-2mm}
\resizebox{0.6\columnwidth}{!}{
\begin{tabular}{|c|c|l|}
\hline
\textbf{Velocity (knots)} & \textbf{Wave Height (m)} & \multicolumn{1}{c|}{\textbf{Wave Frequency (rad)}}         \\ \hline
6                         & 0.30                     & 1.69, 1.83, 1.97, 2.11, 2.25, 2.39, 2.53, 2.67, 2.95       \\ \hline
8                         & 0.30                     & 1.67, 1.83, 1.97, 2.11, 2.25, 2.39, 2.67, 2.95             \\ \hline
8                         & 0.45                     & 1.69, 1.83, 1.97, 2.11, 2.39, 2.53, 2.67, 2.95, 3.23, 3.37 \\ \hline
10                        & 0.30                     & 1.69, 1.83, 1.88,1.97, 2.11, 2.25, 2.39, 2.67, 2.95        \\ \hline
\end{tabular}
\label{modelTestConditions}
}
\end{wraptable}
\FloatBarrier

The towing tank results were converted to full scale using the laws presented in \cite{lloyd1998seakeeping}. Pitch and heave response amplitude operators (RAOs) were determined for each of the vessel velocity and wave height combinations shown in Table~\ref{modelTestConditions}, and converted to dimensionless quantities using the expressions:
    $\text{Pitch Slope}= {|{\theta}_0|}/k_w{\tilde{\zeta}}_0$, and
    $\text{Heave}= |\tau_0|/\tilde{\zeta}_0$,
where $\theta_0$ is the peak to peak pitch in radians, $\tau_0$ is the peak to peak heave in meters, $k_w$ is the wave number, and $\tilde{\zeta}_0$ is the wave amplitude. The dimensionless model test RAOs from measurements for each condition are shown in Figures~\ref{6kn0-3PitchRAO}-\ref{10kn0-3HeaveRAO}. 

\subsubsection{PanShipL Model Validation}\label{sec:pan-model-validation} 
To compute hydrodynamic forces (radiation and damping), the simulation tool, PanShipL, assumes that changes in the submerged geometry at mean forward speed---specifically, the breadth $B$ and draught $T$---are minimal. These assumptions enable the Green functions to be calculated once at the beginning of the simulation and reused at each subsequent time step, thereby improving computational efficiency. However, to compute the wave excitations, the instantaneous submerged geometry of the vessel, which varies over time, is used. The breadth, B, is bounded above by the vessel’s maximum geometric breadth, while the draught, T, can range from zero up to the maximum vertical extent at the centre of gravity (CoG). Then, perturbations about the vessel’s mean speed and heading are assumed to be small.

PanShipL is described as a semi-linear simulation tool because the radiation and diffraction problems are solved in a linearised manner, while the wave excitation and restoring forces are solved in a nonlinear manner using the actual submerged hull geometry under the disturbed incident wave. The disturbed wave is obtained from the pressure at the waterline panels.  
The effect of forward speed on sinkage and trim is determined in calm water at the start of the simulation in conjunction with the resulting submerged hull geometry. A more detailed description of the methods employed by PanShipL can be found in \cite{vanWalree2002} and \cite{vanWalree2009}.

\begin{figure}[!h]
\centering
\includegraphics[width =\textwidth]{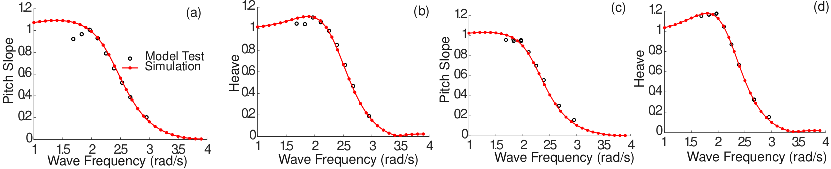}
\caption{6~kn (a) pitch and (b) heave  RAOs for 0.3~m wave. 8~kn {(c)} pitch and {(d)} heave RAOs for 0.3~m wave.}
\label{6kn0-3PitchRAO}
\label{6kn0-3HeaveRAO}
\label{8kn0-3HeaveRAO}
\label{8kn0-3PitchRAO}
\vspace{-2mm}
\end{figure}

\begin{figure}[!h]
\centering
\includegraphics[width =\textwidth]{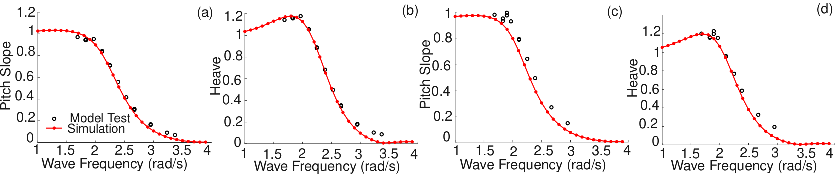}
\caption{8~kn (a) pitch and (b) heave RAO for 0.45~m wave. 10~kn (c) pitch and (d) heave RAOs for 0.3~m wave.}
\label{8kn0-45PitchRAO}
\label{8kn0-45HeaveRAO}
\label{10kn0-3PitchRAO}
\label{10kn0-3HeaveRAO}
\end{figure}

To validate the model, multiple PanShipL simulations were conducted using a RHIB hull model and identical wave conditions to those used in the model test program, as listed in Table~\ref{modelTestConditions}. At the start of each simulation, the input variables were adjusted to align the output more closely with the model test program results. The simulation RAOs presented in Figure~\ref{6kn0-3PitchRAO}-\ref{10kn0-3HeaveRAO}, show that the PanShipL modeling of the RHIB vessel aligns well with the  AMC Towing Tank model test data and provides a reliable means of generating high-fidelity data for evaluating our approach.

\subsection{Experiments with High-Fidelity Data}
The high-fidelity simulation data, generated using the PanShipL numerical modeling tool with the validated RHIB model demonstrated in the Section~\ref{sec:high-fed-data-gen} was used in this section to validate the proposed algorithm. Two sets of data with true parameters $H_s=1.00$ m and $T_z=7.8$ s, under head seas and bow-quartering seas ($3\pi/4$ radians), are used to validate the proposed formulation for possible application in real-time sea operations. In each recording, the vessel moves forward at a speed of $4.11$ m/s, and data is recorded at 10 samples per second for a duration of 15 minutes. the filter initialisation and the construction of the process noise covariance are carried out in the same manner as in the experiments with synthetic data. PanShipL generates displacement and acceleration data for the two vertical vessel motions without sensor noise. However, to account for validation using noisy towing tank data and the presence of frequency components that were used to generate the data but not considered in the estimation process, the following measurement noise covariances are employed: $\mathrm{R}_k = \text{diag}([1.68,\ 1.68,\ 1.68] \times 10^{-5})$ for heave estimation, and $\mathrm{R}_k = \text{diag}([2.38,\ 2.38,\ 2.38] \times 10^{-7})$ for pitch estimation. 
\FloatBarrier
[8])

\subsubsection{Results and Discussion}
The presented formulation and the solutions proposed in Algorithms~\ref{algo1} and \ref{algo2}, along with the KF (known parameters), are used in our evaluation with high-fidelity data. We estimated the unknown vessel parameters, the expected measurements, and the excitations for the two data sets. 

\begin{figure}[h!]%
	\centering
	\begin{subfigure}{.45\columnwidth}
		\includegraphics[width=\columnwidth]{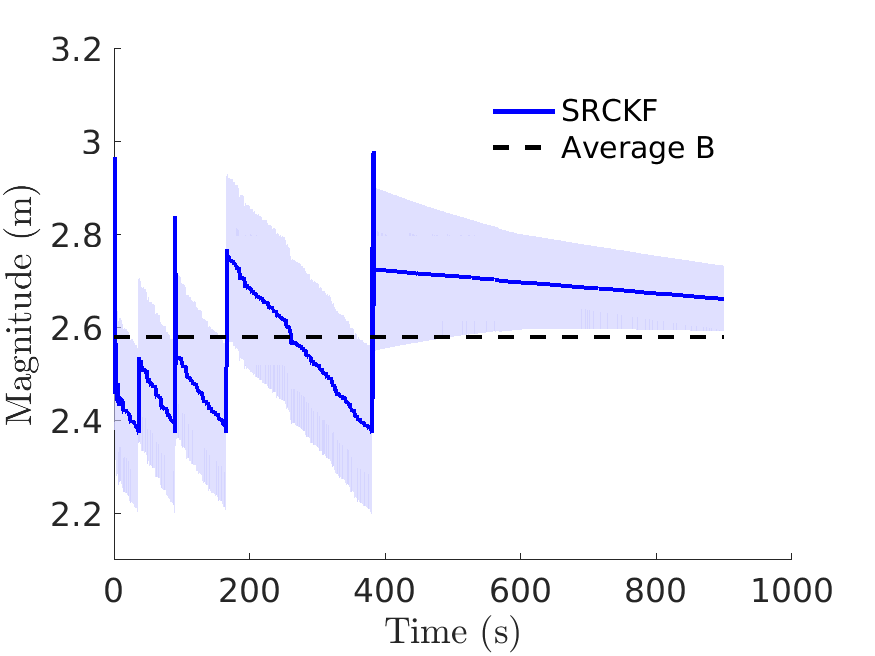}%
		\caption{{Breadth $B$ in meters}.}%
		\label{case2_b_135}%
	\end{subfigure}\hfill%
	\begin{subfigure}{.45\columnwidth}
		\includegraphics[width=\columnwidth]{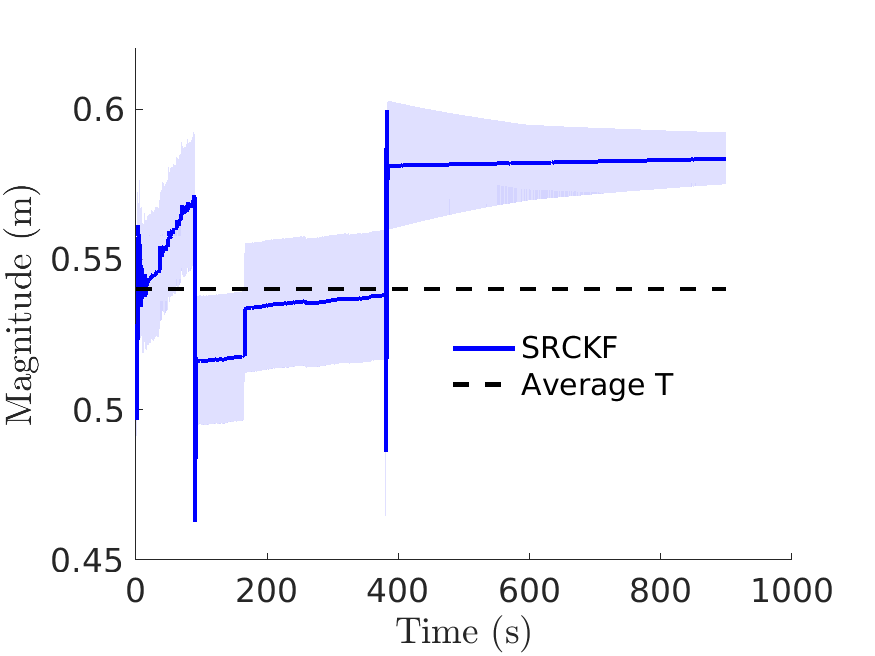}%
		\caption{{Draught $T$ in meters.}}%
		\label{case2_t_135}%
	\end{subfigure}%
    \vspace{-2mm}
	\caption{{Estimation of vessel parameters $B$ and $T$ from high-fidelity data under bow-quartering seas}}
	\label{case2_v}
\end{figure}

\begin{figure}[h!]%
	\centering
	\begin{subfigure}{.49\columnwidth}
		\includegraphics[width=\columnwidth]{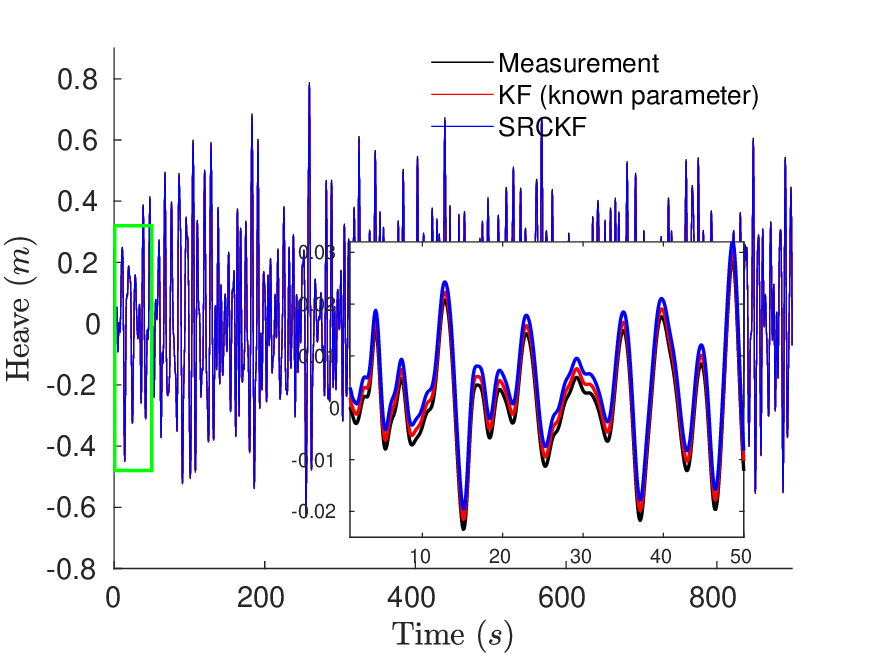}%
		\caption{Heave measurement.}%
		\label{case2_m}%
	\end{subfigure}\hfill%
	\begin{subfigure}{.49\columnwidth}
		\includegraphics[width=\columnwidth]{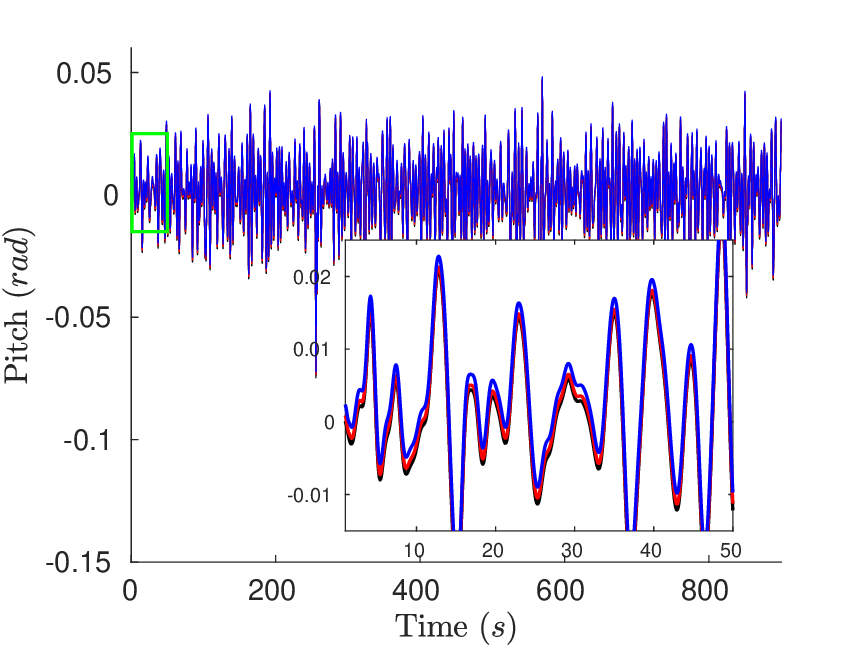}%
		\caption{Pitch measurement.}%
		\label{case2_mp}%
	\end{subfigure}%
    \vspace{-2mm}
	\caption{Estimated expected measurements and their comparison with high-fidelity data under bow-quartering seas.} 
	\label{case2_me}
\end{figure} 
The estimated parameters, $B$ and $T$, are shown in Figure~\ref{case2_v} for the vessel in bow-quartering seas. {The initial sharp variations in these plots reflect the transient behavior of the estimators, which can be attributed to the heave–pitch fusion and the differences between the true dynamics and those assumed in the proposed model.} Importantly, the high-fidelity simulation results in time-varying $B$ and $T$ vessel parameters; therefore, we compare our estimates with the average values of these parameters. Notably, this provides a motivational case study for our formulation, aiming to address the dynamic nature of the vessel parameters and formulate a method to treat them as unknowns to be estimated.

For a given heading, the estimated heave and pitch measurements are plotted along with the actual measurement data in Figures~\ref{case2_m} and \ref{case2_mp}, respectively. It can be observed that the expected measurements from the two methods closely follow the actual measurements from the dataset, thereby confirming the effectiveness of the proposed formulation under these settings and accurately capturing a vessel's motion impacted by sea waves. Although not reported here, similar results are obtained with the heading of $\pi$ radians. 

\begin{figure}[h!]%
	\centering
	\begin{subfigure}{.49\columnwidth}
\includegraphics[width=\columnwidth]{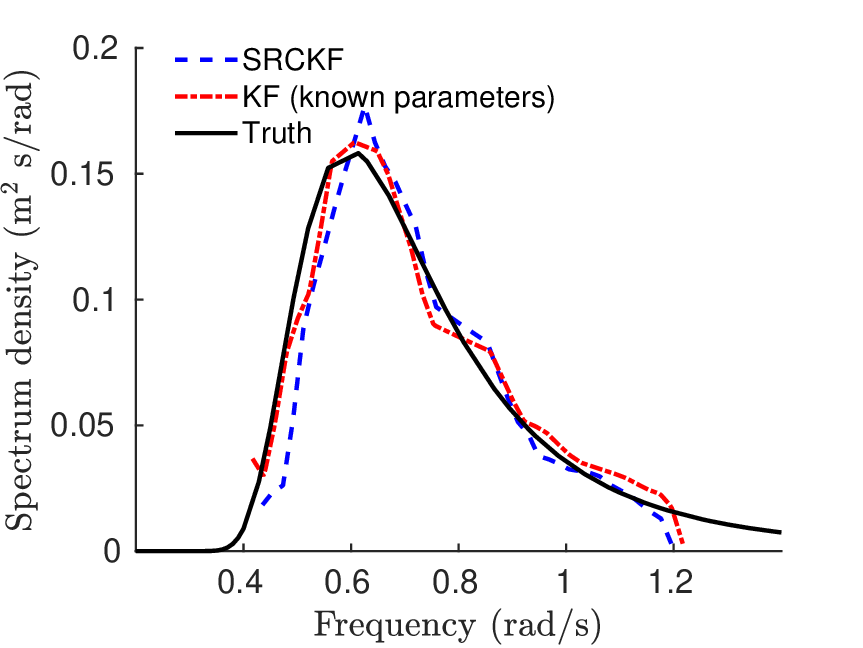}%
		\caption{{Bow-quatering seas}.}%
		\label{case2_m_135}%
	\end{subfigure}\hfill%
	\begin{subfigure}{.49\columnwidth}
		\includegraphics[width=\columnwidth]{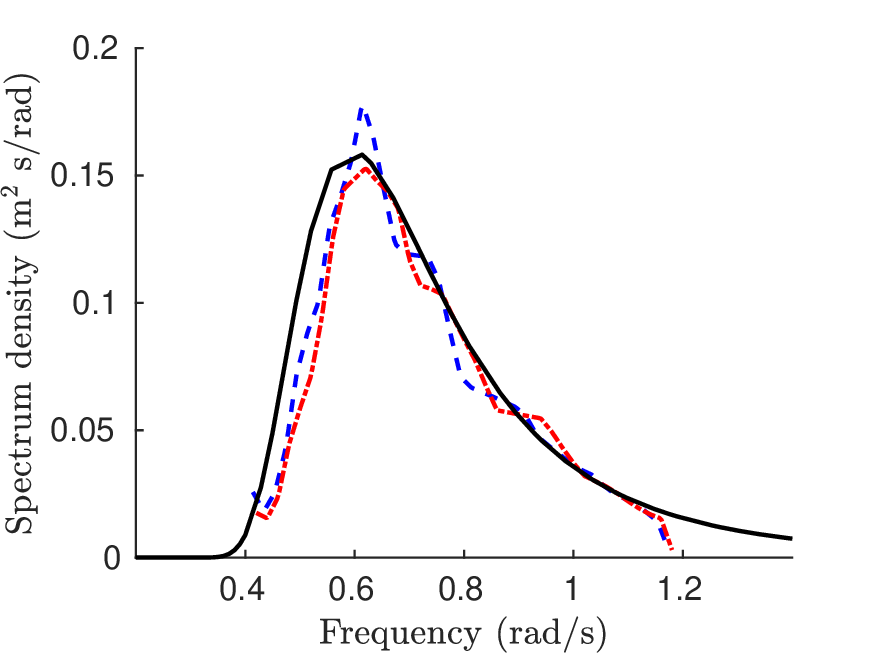}%
		\caption{{Head seas.}}%
		\label{case2_mz_135}%
	\end{subfigure}%
    \vspace{-2mm}
	\caption{{Estimated spectrum from high-fidelity data under different headings.}}
	\label{case3_spec}
\end{figure} 

\begin{table}[h!]
\centering
\caption{Estimated results for the wave spectrum and vessel parameters versus the truth for high-fidelity data.}
\label{tab4}
\vspace{-2mm}
\resizebox{0.8\textwidth}{!}{
\begin{tabular}{|c|c|c|c|c|c|c|c|c|c|c|} 
\cline{2-11}
\multicolumn{1}{c|}{}                          & \multirow{2}{*}{\textbf{Parameters} } & \multirow{2}{*}{\textbf{Truth} } & \multicolumn{2}{c|}{\textbf{SRCKF} }                   & \multicolumn{2}{c|}{\textbf{KF (known)} }              & \multicolumn{2}{c|}{\begin{tabular}[c]{@{}c@{}}\textbf{$|\Delta|$}\\ \textbf{SRCKF}~\end{tabular}} & \multicolumn{2}{c|}{\begin{tabular}[c]{@{}c@{}} \textbf{$|\Delta|$}\\\textbf{KF (known)}\end{tabular}}  \\ 
\hhline{~~~--------|}
\multicolumn{1}{c|}{}                          &                                       &                                  & {\cellcolor[rgb]{0.894,0.894,0.894}}$\pi$   & $3\pi/4$ & {\cellcolor[rgb]{0.894,0.894,0.894}}$\pi$   &$3\pi/4$& {\cellcolor[rgb]{0.894,0.894,0.894}}$\pi$  & $3\pi/4$                                              & {\cellcolor[rgb]{0.894,0.894,0.894}}$\pi$  &$3\pi/4$\\ 
\hhline{|===========|}
\multirow{2}{*}{\textbf{Vessel Parameters}}    & $B$ (m)                               & $2.58$                           & {\cellcolor[rgb]{0.894,0.894,0.894}}$2.68$  & $2.66$   & {\cellcolor[rgb]{0.894,0.894,0.894}}-       & -        & {\cellcolor[rgb]{0.894,0.894,0.894}}4\%      & 3\%                                                     & {\cellcolor[rgb]{0.894,0.894,0.894}}-      & -                                                          \\ 
\hhline{|~----------|}
                                               & $T$ (m)                               & $0.54$                           & {\cellcolor[rgb]{0.894,0.894,0.894}}$0.58$  & $0.58$   & {\cellcolor[rgb]{0.894,0.894,0.894}}-       & -        & {\cellcolor[rgb]{0.894,0.894,0.894}}7\%      & 7\%                                                     & {\cellcolor[rgb]{0.894,0.894,0.894}}-      & -                                                          \\ 
\hline
\multirow{3}{*}{ \textbf{Sea Wave Parameters}} & $H_s$ (m)                             & $1.00$                           & {\cellcolor[rgb]{0.894,0.894,0.894}}$0.94$  & $0.97$   & {\cellcolor[rgb]{0.894,0.894,0.894}}$0.93$  & $0.98$   & {\cellcolor[rgb]{0.894,0.894,0.894}}6\%     & 3\%                                                     & {\cellcolor[rgb]{0.894,0.894,0.894}}7\%      & {2\%}                                                          \\ 
\hhline{|~----------|}
                                               & $T_z$-I (s)                           & $7.80$                           & {\cellcolor[rgb]{0.894,0.894,0.894}}$7.24$  & $7.21$   & {\cellcolor[rgb]{0.894,0.894,0.894}}$7.17$  & $7.31$   & {\cellcolor[rgb]{0.894,0.894,0.894}}7.17\%      & 7.56\%                                                     & {\cellcolor[rgb]{0.894,0.894,0.894}}8.07\%     & 6.30\%                                                          \\ 
\hhline{|~----------|}
                                               & {$T_z$-II (s) }                         & {$7.80$ }                          & {\cellcolor[rgb]{0.894,0.894,0.894}}{$8.56$}  & {$8.46$ }  & {\cellcolor[rgb]{0.894,0.894,0.894}}{$8.43$}  & {$8.44$}   & {\cellcolor[rgb]{0.894,0.894,0.894}}{$9.74\%$}      &{ 8.46\%}                                                     & {\cellcolor[rgb]{0.894,0.894,0.894}}{8.10\%}      & {8.20\%}                                                          \\
\hline
\end{tabular}
}
\end{table}

The estimated wave spectra for the two datasets are shown in Figure~\ref{case3_spec}. The wave parameters computed are presented in Table~\ref{tab4}, along with their corresponding ground truth values for comparison. Here, to understand the difference in estimation accuracy for the high-fidelity simulation of a vessel under a spectrum of irregular waves, we also report the $\Delta\%$ value for each filter compared to the ground truth, defined as the percentage ratio of the absolute error to the ground truth for each parameter. The proposed method achieves an accuracy better than $6\%$ for the significant wave height parameter $H_s$, whereas the baseline KF (with known parameters) yields an accuracy of $7\%$. For $T_z$-I, the proposed method shows an error of $7.56\%$, while the baseline method reports a higher error of $8.07\%$. However, using $T_z$-II, the baseline KF outperforms the proposed method, with a lower error of $8.20\%$. 
Overall, these results highlight the effectiveness of the proposed method in estimating sea spectra and wave parameters, even in the presence of high-fidelity data uncertainty and unknown vessel parameters.

 \section{Conclusions}
We have introduced a new formulation for irregular sea wave estimation employing the conceptual equivalence between mass-spring-damper and wave-vessel interaction, which is adept at handling incomplete knowledge of the wave-vessel transfer function. Building upon our prior research considering regular sea waves, we constructed the system model depicting the interaction between irregular waves and a vessel. Our framework utilises the square-root cubature Kalman filter for an augmented system, enabling a joint estimation of state, unknown wave excitations, and unidentified vessel parameters---specifically, breadth and draught---using measurements from onboard sensors. We offered insights into the statistics of deriving process noise for such a problem formulation, which is vital for implementing the proposed estimation algorithm. Then, utilising synthetically generated measurements from a model-scale test program, we demonstrated the effectiveness of our formulation in achieving performance comparable to conventional methods relying on accurate parameters of vessel dynamics parameter. Furthermore, to test the validity and applicability of the proposed formulation in practical operational conditions, we use high-fidelity datasets and demonstrate that the estimated wave properties are comparable to their true values when vessel parameters are unknown and changing. Overall, our results indicate that the algorithm we investigated is effective and presents an important step towards addressing a practical problem in sea estate estimation.
More generally, we believe that the results presented in this study can be helpful for many maritime applications as well as for improving meteorological models by fusing additional sources of wave observation data.

{The present study employs two-dimensional unidirectional long-crested wave fields to model wave–vessel interactions under different vessel heading directions. While such two-dimensional random wave fields can capture aspects of ocean conditions, such as swell seas and deep ocean waves, ocean environments are often more complex, including three-dimensional short-crested wave fields with broad directional spreading. Therefore, future work will address short-crested waves and investigate incorporating the vessel length as an additional unknown parameter. To facilitate this extension, we expect the need to consider additional motion components beyond vertical motions to capture more diverse information, thereby facilitating the estimation problem. Our work provides a solid theoretical foundation to investigate these problems further.}
%
 \begingroup
\footnotesize
  \bibliographystyle{elsarticle-harv} 
 \bibliography{references}
\endgroup
\end{document}